\documentclass{jfm}

\usepackage{graphicx}
\usepackage{natbib}

\ifCUPmtlplainloaded \else
  \checkfont{eurm10}
  \iffontfound
    \IfFileExists{upmath.sty}
      {\typeout{^^JFound AMS Euler Roman fonts on the system,
                   using the 'upmath' package.^^J}%
       \usepackage{upmath}}
      {\typeout{^^JFound AMS Euler Roman fonts on the system, but you
                   dont seem to have the}%
       \typeout{'upmath' package installed. JFM.cls can take advantage
                 of these fonts,^^Jif you use 'upmath' package.^^J}%
      }
  \else
  \fi
\fi

\ifCUPmtlplainloaded \else
  \checkfont{msam10}
  \iffontfound
    \IfFileExists{amssymb.sty}
      {\typeout{^^JFound AMS Symbol fonts on the system, using the
                'amssymb' package.^^J}%
       \usepackage{amssymb}%
         
         \let\geq=\geqslant
      }{}
  \fi
\fi

\ifCUPmtlplainloaded \else
  \IfFileExists{amsbsy.sty}
    {\typeout{^^JFound the 'amsbsy' package on the system, using it.^^J}%
     \usepackage{amsbsy}}
    {\providecommand\boldsymbol[1]{\mbox{\boldmath $##1$}}}
\fi

\newcommand\bb{\begin{eqnarray}}
\newcommand\ee{\end{eqnarray}}

\newcommand\bs{\boldsymbol}

\newsavebox{\astrutbox}
\sbox{\astrutbox}{\rule[-5pt]{0pt}{20pt}}

\usepackage{color}
\newcommand{\red}{\color{red}}

\title[Turbulent windprint on a liquid surface]{Turbulent windprint on a liquid surface}

\author[S. Perrard \textit{et al.}]%
{St\'ephane Perrard$^{1,2}$, Adri\'an Lozano-Dur\'an$^{3}$, Marc Rabaud$^{1}$, \\Michael Benzaquen$^{2}$ and Fr\'ed\'eric Moisy$^{1}$}

\affiliation{$^1$FAST, CNRS, Universit\'e Paris-Sud, Universit\'e Paris-Saclay, 91405 Orsay, France\\[\affilskip]
$^2$LadHyX,  UMR CNRS 7646, Ecole polytechnique, 91128 Palaiseau, France\\[\affilskip]
$^3$Center for Turbulence Research Stanford University, Stanford, CA 94305, USA}

\pubyear{2018}
\date{\today; revised ?; accepted ?. - To be entered by editorial office}
\begin{document}

\maketitle

\begin{abstract}
We investigate the effect of a light turbulent wind on a liquid surface, below the onset of wave generation. In that regime, the liquid surface is populated by small disorganised deformations elongated in the streamwise direction. Formally identified recently by Paquier \textit{et al.} (2015), the deformations that occur below the wave onset were named wrinkles. We provide here a theoretical framework for this regime, using the viscous response of a free liquid surface submitted to arbitrary normal and tangential interfacial stresses at its upper boundary. We relate the spatio-temporal spectrum of the surface deformations to that of the applied interfacial pressure and shear stress fluctuations. For that, we evaluate the spatio-temporal statistics of the turbulent forcing using Direct Numerical Simulation of a turbulent channel flow, assuming no coupling between the air and the liquid flows. Combining theory and numerical simulation, we obtain synthetic wrinkles fields that reproduce the experimental observations. We show that the wrinkles are a multi-scale superposition of random wakes generated by the turbulent fluctuations. They result mainly from the nearly isotropic pressure fluctuations generated in the boundary layer, rather than from the elongated shear stress fluctuations. The wrinkle regime described in this paper naturally arises as the viscous-saturated asymptotic of the inviscid growth theory of \citet{Phillips_1957}. We finally discuss the possible relation between wrinkles and the onset of regular quasi-monochromatic waves at larger wind velocity. Experiments indicate that the onset of regular waves increases with liquid viscosity. Our theory suggests that regular waves are triggered when the wrinkle amplitude reaches a fraction of the viscous sublayer thickness. This implies that the turbulent fluctuations near the onset may play a key role in the triggering of exponential wave growth.
\end{abstract}


\section{Introduction}

A mirror like liquid surface is quite rare to observe in outdoor conditions. The smallest breeze already perturbs the surface of water well below the onset of wave formation, as first described by J.S. \citet{Russell_1844}. Since Russel's work, this weak deformation regime below the wave onset has been often reported~\citep{Keulegan_1951,Gottifredi_1970,Kahma_1988,Zhang_1995,Caulliez_2008} but its precise spatio-temporal properties have been measured only recently by~\citet{Paquier_2015} who named these deformations \textit{wrinkles}. Characterised by randomly distributed, elongated structures in the streamwise direction, the apparent disordered aspect of the wrinkles were interpreted qualitatively as a signature of the air turbulence on the liquid surface. 
An important feature of wrinkles is their dependence on the liquid viscosity $\nu_\ell$. An empirical relation was inferred for the mean square amplitude~\citep{Paquier_2016}, 
\bb
\overline{\zeta^2} \propto \nu_{\ell}^{-1} u^{*3} ,
\label{eq_paquier}
\ee
where $u^*$ is the friction velocity of the air~\citep{Schlichting_2000} and $\overline{\zeta^2}$ is averaged over space and time. To the best of our knowledge, this empirical expression has not been explained theoretically.

Since wrinkles are systematically found even at moderate wind, they naturally represent a base state on which regular (quasi-monochromatic) waves propagating in the wind direction may grow. The influence of this initial perturbed surface state on the wave onset can be indirectly analysed by varying the liquid viscosity. Indeed, in the general understanding of the physical origin of the wind wave onset, the dependency in liquid viscosity is still lacking~\citep{Sullivan_2010}. While it has been observed experimentally that the liquid viscosity influences the wave onset \citep{Francis_1956,Kahma_1988,Veron_2001,Paquier_2016}, the explicit dependency has not been captured by models that include viscous effects~\citep{Lindsay_1984,Funada_2001,Kim_2011}.
  
 In the literature of wind wave generation, two families of models can be identified. They involve either stability analysis of the mean wind profile or turbulent fluctuations. The branching goes back to the two seminal papers of \citet{Miles_1957} and \citet{Phillips_1957} who proposed each a mechanism for wind wave generation. On the one hand, the stability analysis based on the mean wind profile in Miles models~\citep{Miles_1993} or on the general Orr-Sommerfeld equation~(see, e.g., \citealp{manneville2010instabilities}) ignores the turbulent fluctuations. Linear stability analysis predicts an onset above which the wave amplitude grows exponentially in time. A qualitative agreement with Miles theory has been obtained in laboratory experiments~\citep{Kawai_1979,Veron_2001}. However, a quantitative agreement of refined Miles models~\citep{Janssen_2004} with experimental data is still lacking both in laboratory conditions~\citep{Plant_1982,Liberzon_2011} and in outdoor conditions~\citep{Sullivan_2010}. On the other hand, Phillips, following \citet{Eckart_1953}, analysed \textcolor{black}the effect of random pressure fluctuations at the surface of an inviscid liquid. He considered a resonance mechanism between the surface displacement $\zeta$ and the pressure fluctuations $p$. He derived an expression for the mean square surface displacement of the form
\bb
\overline{\zeta^2} = \frac{1}{\rho_\ell^2} \frac{\overline{p^2} \, t}{2\sqrt{2} U_\mathrm{c} g} ,
\label{eq_phillips}
\ee
where $\overline{p^2}$ is the mean square pressure fluctuation averaged over space, $\rho_\ell$ the liquid density, $U_\mathrm{ c}$ the typical convection speed of the turbulent structures and $g$ the acceleration of gravity. Phillips theory yields a linear growth for the wave energy. Contrary to Miles theory, Phillips theory has not been extensively tested. Experimentally, algebraic temporal growth for the surface deformation has been observed recently by~\citet{Zavadasky_2017}. Recent direct numerical simulations of two phase shear flows have also observed a regime of algebraic growth in time~\citep{Lin_2008,Zonta_2015} that may be attributed to Phillips mechanism. However, a quantitative agreement with Phillips theory has not been reported to date.

Phillips formalism may apply to the wrinkle regime, as the air flow is already turbulent for wind below the wave threshold, so that a resonance between pressure fluctuations and surface waves could already occur at low wind speed. Equation~(\ref{eq_phillips}) provides a theoretical expression for $\zeta$ that depends on time and is independent of viscosity, whereas Eq.~(\ref{eq_paquier}) provides an empirical relation for $\zeta$ that depends on viscosity and is independent of time. These two equations are therefore far apart elements of the wrinkle puzzle. The aim of this paper is to provide a theoretical framework for the wrinkle regime that reconciles Eqs.~(\ref{eq_paquier}) and (\ref{eq_phillips}).

A quantitative description of the wrinkle regime, even though it does not involve any instability mechanism, is challenging for mainly two reasons. First, the formalism must describe the response of a viscous liquid to an arbitrary forcing both in time and in space. For an impulsive forcing of arbitrary shape one can follow the approach of \citet{Miles_1968}, who revisited the Cauchy-Poisson problem~\citep{Lamb_1995} for a viscous liquid. The response to a continuous perturbation in time can also be computed using the same formalism by linear superposition. However, the Fourier-Laplace transform formalism used by Miles limits the analytical feasibility to asymptotic solutions. \textcolor{black}{For the specific case of a pressure source travelling at constant velocity, the wave pattern generated at the surface of an inviscid liquid was computed by~\cite{havelock1919}. This classical problem, which provides a simplified description of the far-field wake behind a ship, has been recently revisited, for an inviscid~\citep{Raphael_1996,Rabaud_2013,Darmon_2014} or a viscous~\citep{Richard_1999} fluid.} To the best of our knowledge, no such Havelock-like formulation is yet available for an arbitrary forcing pattern both in time and in space. Moreover, the Havelock formulation applies only for a pressure disturbance, whereas both pressure and shear stress act on the surface of a viscous liquid.

The second main difficulty arises from the modelling of the turbulence in the air boundary layer. \textcolor{black}{Of particular interest for the wave generation problem is the slow dynamics and the long-range correlations of the pressure and shear stress fluctuations in the boundary layer. Their main statistical properties (characteristic size, mean convection velocity) have been measured since the 60's, e.g. by \cite{Willmarth_1962} and \cite{Corcos_1963}; see~\cite{Robinson_1991} for a comprehensive review  prior to the development of numerical simulations. These early quantitative measurements, obtained from single-point probes, however, could not provide a full spatio-temporal characterisation of the turbulent fluctuations in a boundary layer}.

In recent years, in-depth knowledge has been gained from Direct Numerical Simulations (DNS), both for developing turbulent boundary layers and for fully developed turbulent channel flows \citep{Choi_1990,Moser_1999,Jimenez_2004,Jimenez_2008}. The maximum turbulent Reynolds number reached in the most advanced simulations, Re$_\tau \simeq 4000 - 8000$~\citep{lozano2014time,Lee_JFM_2015,Yamamoto_PRF_2018}, is comfortably larger than the relevant values for the wind-wave generation problem (Re$_\tau \simeq 100-1000$). Such data are highly valuable for the study of the wrinkle regime, as the air turbulence can be considered as essentially unaffected by the wave motion. Indeed, the typical surface displacement amplitude is much smaller than the viscous sublayer thickness in the air flow. This is precisely the regime explored in this paper.   The situation is different for larger amplitude, close to the wave onset, for which a feedback of the wave motion on the air turbulence is expected, requiring a full two-phase flow approach. This approach, much more  demanding computationally speaking, has been investigated recently~\citep{belcher1998turbulent,Kudryavtsev_2002,Lin_2008,Druzhinin_2012,Kudryavtsev_2014,sajjadi2017wave}.
Because of the high computational cost, the range of physical parameters covered by these studies remains limited, and no scaling relation has yet been obtained for the wave onset.
	
\textcolor{black}{An important feature in the generation of surface waves by the wind is the unavoidable presence of a mean advecting current in the liquid, which may significantly affect the wave dynamics~\citep{Peregrine_1976,Kirby_1990,Ellingsen_2017}. The growth and saturation of a drift velocity in the liquid and wave growth are two closely intertwined processes~\citep{Banner_1998,Melville_1998,Veron_2001}, and the stability of the flow depends on the resulting velocity profiles both in the air and in the liquid~\citep{Miles_1993}. However, in the wrinkle regime explored by~\cite{Paquier_2015}, the drift velocity is bounded to a few cm/s by the finite channel depth and the large viscosity of the fluid. Such drift velocity remains small compared to both air velocity and phase velocity of the wrinkles, and is neglected in the present study.}

In this paper we focus on a quantitative description of the wrinkle regime, below the wind wave onset. For that purpose, we combine analytical calculations of the viscous response of the liquid to an arbitrary forcing in a statistically stationary state with DNS data of a turbulent channel flow. Based on careful dimensional analysis and experimental data of the wrinkle regime, we show that below the wave threshold the evolution of the turbulent air boundary layer can be decoupled from the liquid response. We thus circumvent the difficulty to simulate a full turbulent two-phase flow by considering a turbulent channel flow with rigid walls and no-slip boundary conditions. Doing so, we greatly simplify the numerical setup to focus on the linear, passive response of the liquid phase. We show that the wrinkle statistics computed from our model are in good agreement with the experimental data of~\citet{Paquier_2015,Paquier_2016}. Finally, we show that the onset of regular waves may correspond to the breakdown of the regime of linear passive response of the liquid surface, and propose an empirical criterion based on surface roughness originating from the wrinkle amplitude.

The paper is organized as follows. Section 2 provides a general dimensional analysis of the surface deformation generated by a turbulent boundary layer. In section 3 we derive the key equation of the paper, that establishes the link between the Fourier components of the normal and tangential shear stresses applied at the air-liquid interface and the Fourier components of the surface displacement. Section 4 presents the direct numerical simulations used to compute the normal and tangential stresses.  Section 5 combines the equation for the surface displacement in Fourier space and the output of the DNS to compute the wrinkle properties. It provides a quantitative comparison with the experiments of~\citet{Paquier_2016}. Section 6 finally bridges the gap between the wrinkle regime governed by viscosity and the inviscid resonant theory of \citet{Phillips_1957}.

\section{Wrinkle regime: dimensional analysis and experimental set up}
\subsection{Dimensional Analysis}

\begin{figure}
\centering
\includegraphics[width=0.75 \columnwidth]{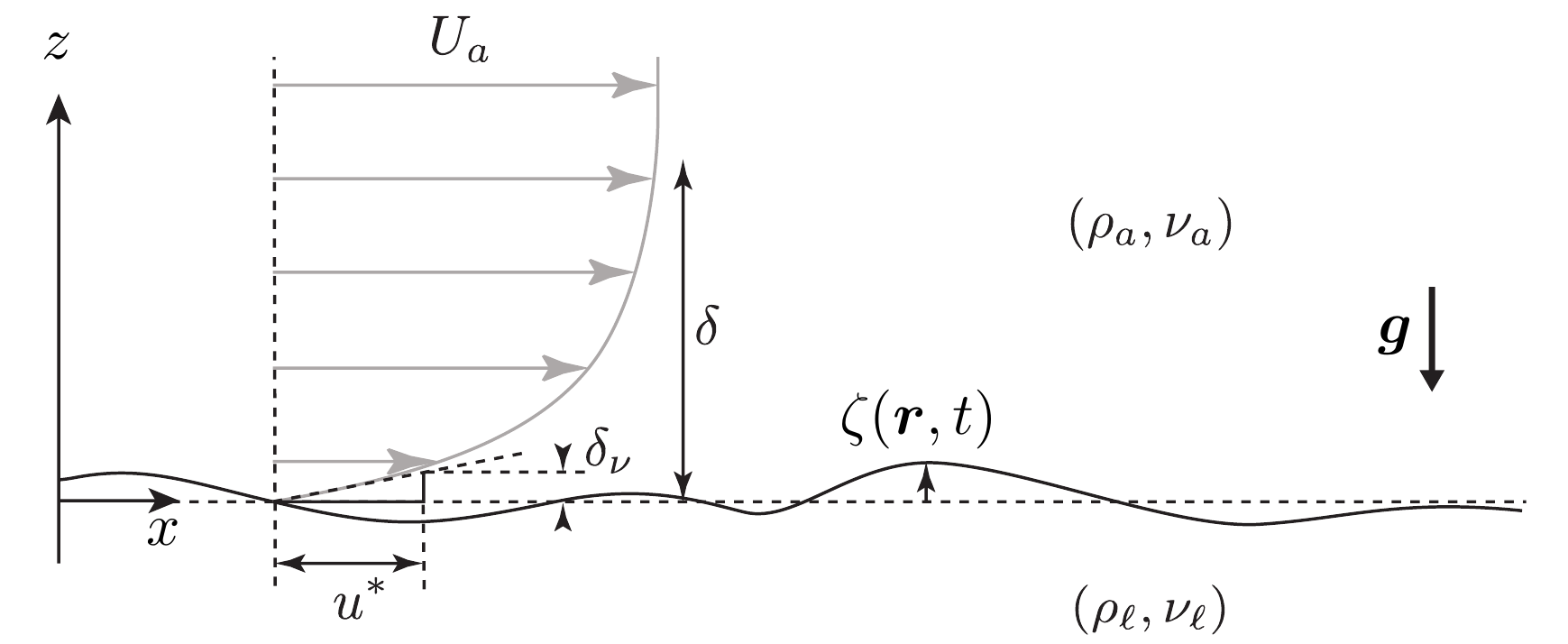}
\caption{Sketch of the velocity profile of a turbulent wind blowing on a viscous liquid. The turbulent boundary layer is characterised by the outer layer thickness $\delta$ and the viscous sublayer thickness $\delta_\nu = \nu_\mathrm{a}/u^*$. The friction velocity $u^*$ is defined from the mean shear stress $\tau_\mathrm{a} = \rho_\mathrm{a} {u^*}^2$ at the wall. The liquid response may depend on the air/liquid densities $\rho_\mathrm{a},\rho_\ell$, the kinematic viscosities $\nu_\mathrm{a},\nu_\ell$, as well as $\delta,u^*$, the gravity $g$ and the surface tension $\gamma$.}
\label{fig0}
\end{figure}

We first discuss here the dependence of the surface displacement $\zeta$ on the relevant physical parameters using dimensional analysis. In the statistically steady state, the characteristic amplitude of the surface displacement averaged over space and time $\zeta_\mathrm{rms} = (\overline{\zeta^2} )^{1/2}$ in response to a turbulent wind is expected to depend on numerous parameters that characterise both the turbulent flow in the air and the liquid properties. The geometry and the relevant parameters of both phases are sketched in figure~\ref{fig0}. The turbulent air boundary layer is characterised by the air density $\rho_\mathrm{a}$, the kinematic viscosity $\nu_\mathrm{a}$, the friction velocity $u^* = \sqrt{\tau_\mathrm{a} / \rho_\mathrm{a}}$ (where $\tau_\mathrm{a}$ is the mean shear stress at the interface), and the boundary layer thickness $\delta$. In a developing boundary layer, the thickness $\delta$ is function of the streamwise distance $x$, usually called {\it fetch} in the wind-wave generation problem, whereas it is constant in a fully developed channel flow. The spatial variation of a developing boundary layer is usually small ($\mathrm{d} \delta / \mathrm{d} x \ll 1$), so we can simply consider the boundary layer thickness $\delta$ as constant. The liquid flow is characterised by the liquid density $\rho_\ell$, kinematic viscosity $\nu_\ell$, acceleration of gravity $g$, surface tension $\gamma$, and liquid depth $h$.

Under these hypotheses, the amplitude of the surface displacement writes $\zeta_\mathrm{rms}= f(\rho_\mathrm{a},\rho_\ell,\nu_\mathrm{a},\nu_\ell,\delta,u^*,g, \gamma, h)$. According to Buckingham's $\pi$ theorem, the dimensionless surface displacement $\zeta_\mathrm{rms}/\delta$ can be expressed as a function of six independent dimensionless numbers. Choosing a set of dimensionless numbers that decouple the influence of the friction velocity $u^*$ and of the length scale $\delta$, it writes
\bb
\frac{\zeta_\mathrm{rms}}{\delta} = f_1 \left (\frac{\rho_\mathrm{a}}{\rho_\ell}, \frac{g \delta^3}{\nu_\ell^2}, \frac{{u^*}^3}{g \nu_\ell}, \textrm{Re}_\delta, \textrm{Bo}, \frac{h}{\delta} \right),
\ee
where $f_1$ is a dimensionless function, $\textrm{Re}_\delta = u^* \delta /\nu_\mathrm{a}$ is the turbulent Reynolds number characterising the boundary layer, and Bo~$= \delta / \ell_c$ is the Bond number (with $\ell_c = \sqrt{\gamma / \rho_\ell g}$ the capillary length).  \textcolor{black}{The relative depth $h/\delta$ becomes relevant for surface wave propagation in the shallow water regime. We are mostly interested here in the the deep water regime (one has $h/\delta \simeq 1.2$ in the experiment, see section 2.2), so the importance of this parameter is marginal in the following.}

The function $f_1$ can be further specified using additional physical arguments. In the static case and without surface tension, the density ratio $\rho_\mathrm{a}/\rho_\ell$ sets the surface displacement amplitude as can be seen from a simple pressure balance: the gravity pressure scales as $\rho_\ell g \zeta$ and the pressure fluctuations in the air phase scales as $\rho_\mathrm{a} {u^*}^2$. Adding the surface tension introduces a dependency in Bond number but it does not modify the scaling of $\zeta_\mathrm{rms}$ in $\rho_\mathrm{a}/\rho_\ell$. In the dynamical case with negligible effect of gravity in the air phase, and in the limit of linear equation of motion, the displacement still scales as $\rho_\mathrm{a}/\rho_\ell$ as can be seen from the continuity of normal stresses at liquid-air interface (a proper justification is given in Section 3). It implies
\bb
\frac{\zeta_\mathrm{rms}}\delta = \frac{\rho_\mathrm{a} }{\rho_\ell} f_2 \left (\frac{g \delta^3}{\nu_\ell^2}, \frac{{u^*}^3}{g \nu_\ell}, \textrm{Re}_\delta , \textrm{Bo}, \frac{h}{\delta} \right) .
\ee
The dimensionless number $g \delta^3/\nu_\ell^2$ characterises the surface response of a viscous liquid to an initial perturbation, as analysed by~\citet{Miles_1968}. The dependency in $g \delta^3/\nu_\ell^2$ can be neglected for the following reason. We first rewrite $g \delta^3/\nu_\ell^2$ as $(\delta / \ell_\nu)^3$, with $\ell_\nu = g^{-1/3} {\nu_\ell}^{2/3}$ the viscous length. This length $\ell_\nu$ was identified by \citet{Miles_1968} as the relevant length scale to classify the surface deformation regimes. For a gravity wave of wavenumber $\delta^{-1}$ and angular frequency $\omega = \sqrt{g/\delta}$, the dimensionless damping factor defined as $\theta = \nu_\ell/(\omega \delta^2)$ is given by $\theta = (\ell_\nu/\delta)^{3/2}$. The separation between the propagating wave regime and the over-damped regime occurs at a finite value of $\theta$, namely $\theta_c = 1.31$~\citep{Leblond_1987}. Hence for gravity waves $\ell_\nu$ separates the regime of propagating waves ($\theta < \theta_c$) from the over-damped regime $(\theta > \theta_c)$. In the range of liquid viscosity for which wrinkles are observed, $\nu_\ell \simeq 1-10^3 \nu_{\rm water}$, the viscous length $\ell_\nu$ is in the range 50~$\mu$m $-$ 5~mm. Natural and laboratory flows characterised by a forcing scale $\delta >$1~cm and for moderate viscosity $\nu_\ell <10^3 \nu_\mathrm{water}$ therefore fall in the regime $\delta \gg \ell_\nu$ in which gravity dominates over viscous diffusion. Surface deformations will therefore not be significantly diffused horizontally by viscous effects, and we may thus neglect the influence of $g \delta^3/\nu_\ell^2$ on the surface deformation.

Even though the dimensionless number $g \delta^3/\nu_\ell^2$ is non relevant in the propagating regime, viscosity cannot be neglected because its cumulative effect eventually balances the input forcing~\citep{Miles_1968}. The displacement $\zeta_\mathrm{rms}$ may therefore depend on ${u^*}^3/(g \nu_\ell)$, and one is left with
\bb
\frac{\zeta_\mathrm{rms}}{\delta} = \frac{\rho_\mathrm{a}}{\rho_\ell} f_3 \left (\frac{{u^*}^3}{g \nu_\ell}, \textrm{Re}_\delta , \textrm{Bo}, \frac{h}{\delta} \right) .
\label{eq:f3}
\ee
{\textcolor{black}Finally, from a balance between the energy flux from the turbulent boundary layer in the air and the viscous loss in the liquid, it is possible to infer the dependence of $\zeta$ in the liquid viscosity $\nu_\ell$. Neglecting surface tension and finite depth effects (large $\mathrm{Bo}$ and large $h/\delta$), the potential energy $e$ per unit surface of a deformation of amplitude $\zeta$ is of order $\rho_\ell g \zeta^2$. If this fluctuation is in the wave regime (weak viscous attenuation), its kinetic energy per unit surface is also of order $e$.  We assume here that the liquid surface is mostly sensitive to the largest scales of the turbulent fluctuations, governed by the boundary layer thickness $\delta$ (we provide support to this key assumption in section 5).  Consider a vertical velocity fluctuation of order $u^*$ over horizontal extent of order $\delta$, corresponding to a pressure fluctuation of order $\rho_a u^{*2}$, pushing or sucking the liquid surface at a velocity $\dot \zeta$. Conservation of vertical momentum in this inelastic process implies $\rho_a u^* \simeq \rho_\ell \dot \zeta$. The power per unit surface transferred to the liquid, $\rho_a u^{*2} \dot \zeta$, therefore writes $(\rho_a^2 / \rho_\ell) u^{*3}$. In a statistically steady state, this power must be balanced by the energy loss by viscous diffusion in the liquid, $e/\tau$, with $\tau \simeq \delta^2 / \nu_\ell$ the viscous time scale, yielding $\zeta \simeq \delta (\rho_a / \rho_\ell) (u^{*3}/g\nu_\ell)^{1/2}$. This suggests writing Eq.~(\ref{eq:f3})  in the form\footnote{This dimensional form was anticipated in \cite{Paquier_2016}, but with a wrong exponent in $(\rho_a / \rho_\ell)$ and a different definition for the forcing scale $\delta$.}
\bb
\frac{\zeta_\mathrm{rms}}{\delta} = \frac{\rho_\mathrm{a}}{\rho_\ell} \left (\frac{{u^*}^3}{g \nu_\ell} \right)^{1/2} f_4 \left (\textrm{Re}_\delta, \textrm{Bo}, \frac{h}{\delta} \right).
\label{eq:f4}
\ee
The aim of this paper is to provide a solid mathematical ground to this scaling using viscous surface waves theory and an appropriate evaluation of the surface response in Fourier space following a route similar to \citet{Phillips_1957}.

In the following, the full dependency of $\zeta/\delta$ with respect to the Bond number and liquid depth is not explored, as the experimental data on the wrinkle regime are available only for one value of Bo and $h/\delta$ (see section 2.2 for details). The remaining dependency of $\zeta/\delta$ in $\textrm{Re}_\delta$ is subtle and rises the question of the relevant forcing scale in this problem. The above qualitative argument assumed that the forcing acts at the scale of the boundary-layer thickness $\delta$, whereas the spectrum of the stress fluctuations in a turbulent boundary layer spreads in the range from $\delta$ down to the viscous sublayer thickness $\delta_\nu = \mathrm{Re}_\delta^{-1} \delta$. The non-trivial dependence in $\textrm{Re}_\delta$ therefore depends on the exact forcing spectrum and the nature of the surface response, and will be characterised in section 5 using DNS data. The key result of the paper is that, in the range of $\textrm{Re}_\delta$ relevant for the experiments, the function $f_4$ is essentially constant, which is consistent with the empirical scaling (\ref{eq_paquier}).}

\subsection{Experimental details}

Although the theory derived in this paper is general, quantitative comparison  in the following  is provided with the only  available experimental data of \cite{Paquier_2015,Paquier_2016}. We briefly provide here some details about the experiment, \textcolor{black}{and summarise the relevant non-dimensional numbers in table~1.}

The experimental set up consists in a rectangular tank filled with liquid, fitted to the bottom of a wind-tunnel of rectangular cross-section. The tank is of length $L=1.5$~m, width $296$~mm, and depth $h=35$~mm, and the channel height is $105$~mm. Air (density $\rho_a \simeq 1.2$~kg~m$^{-3}$, kinematic viscosity $\nu_a \simeq 15 \times 10^{-6}$~m$^2$~s$^{-1}$) is injected at a mean velocity $U_a$ in the range $1-10$~m~s$^{-1}$. The velocity profile in the air is close to that of a classical turbulent boundary layer developing over a no-slip flat wall, at least in the wrinkle regime. The boundary-layer thickness $\delta$, defined as the distance from the surface at which the mean velocity is $0.99 U_a$, is $\delta \simeq 13$~mm at $x=0$, and increases linearly along the tank, with $d \delta / dx \simeq 0.02$. At the fetch $x$ at which the measurements are performed, the local boundary layer thickness is $\delta \simeq 30$~mm. The friction velocity $u^*$, determined from measurement of the surface drift velocity using stress continuity, is approximately $u^* \simeq 0.05 U_a$.

The liquid viscosity is varied in a wide range, $\nu_\ell = 0.9 - 560 \times 10^{-6}$~m$^2$~s$^{-1}$, using mixtures of glycerol and water (for low $\nu_\ell$) or glucose syrup and water (for large $\nu_\ell$). The liquid density $\rho_\ell$ of the mixtures is in the range $1.0-1.36 \times 10^3$ kg~m$^{-3}$. The surface tension is $\gamma \simeq 60$~mN~m$^{-1}$, and the capillary length $\ell_c= \sqrt{\gamma / \rho_\ell g}$ is approximately 2.2~mm.

For a laminar flow in the liquid, the drift velocity at the surface is given by the continuity of the shear stress, namely $U_s = \rho_a u^{*2} h / (4 \rho_\ell \nu_\ell)$. For liquid viscosity $\nu_\ell > 4 \nu_{\rm water}$, the maximum drift velocity in the wrinkle regime is $U_s \simeq 10$~cm/s (for a wind velocity $U_a \simeq 4.5$~m/s), for which the flow in the liquid remains essentially laminar ($U_s h / \nu_\ell < 10^3$). For the largest liquid viscosity, $\nu_\ell \simeq 600 \nu_{\rm water}$, the surface drift does not exceed 1~mm/s even at the largest wind velocity, and can be safely neglected. However, for water and liquids of viscosity up to $\simeq 4 \nu_{\rm water}$, the drift velocity is significant and the flow in the liquid is no longer laminar. For the sake of simplicity, the effects of the drift velocity and of turbulence in the liquid are not considered in the present paper.

The instantaneous surface deformation fields $\zeta(\bs r,t)$ are measured using Free-Surface Synthetic Schlieren~\citep{Moisy_2009}. This optical method is based on the analysis of the refracted image of a pattern visualized through the interface. The field of view is $390 \times 280$~mm. The horizontal resolution is 3~mm, and the vertical resolution is 0.6~$\mu$m. The rms wrinkle amplitude is in the range 1-10~$\mu$m ($\zeta_{rms} / \delta \simeq 3 \times 10^{-5} - 3 \times 10^{-4}$).

\begin{table}
\begin{center}
\begin{tabular}{p{3cm} p{3cm} }  
Re$_\delta = u^* \delta / \nu_a$ &  $10^2-10^3$  \\
Bo $= \delta/\ell_c$     &  14 \\
$h/\delta$               &  1.2 \\
$\rho_a/\rho_\ell$       &  $0.9 \times 10^{-3} - 1.2 \times 10^{-3} $ \\
$u^{*3}/g \nu_\ell$      &  $2 \times 10^{-2} - 2 \times 10^2$ \\
$g \delta^3/\nu_\ell^2$  &  $ 8 \times 10^2 - 3 \times 10^8$
\label{tableNDN}
\end{tabular}
\end{center}
\caption{\textcolor{black}{Set of non-dimensional numbers in the experiments \citep{Paquier_2015,Paquier_2016}.}}
\end{table}

\section{Derivation of the spectral theory}

In this section we derive an expression in Fourier space relating the surface displacement to an arbitrary pressure and shear stress fields applied at a liquid interface from the upper gas phase. 

\subsection{Assumptions}

The calculation is made under the following assumptions:
\begin{itemize}
\item[(a)] The gas density is small compared to the liquid density, $\rho_\mathrm{a} \ll \rho_\ell$.
\item[(b)] The slope of the surface displacement stays small at all time ($|\nabla \zeta | \ll 1$).
\item[(c)] The flow in the liquid is laminar.
\item[(d)] The drift velocity in the liquid is negligible compared to the convection speed of the turbulent structures and the typical phase velocity of the wrinkles. 
\item[(e)] The liquid layer is of infinite depth. 
\end{itemize}

Assumptions (a)-(c) are fundamental hypotheses of our theoretical approach. \textcolor{black}{The drift current (assumption (d)) strongly affects the propagation of surface wave in oceanographic context~\citep{Peregrine_1976}, and correction to the dispersion relation has to be taken into account~\citep{Kirby_1990,Ellingsen_2017}. However, in our experimental range of parameters (see section 2.2), the drift current stays within few per cent of the convection speed of the turbulent structures and the typical phase velocity of the wrinkles, so it can be neglected.} The finite depth correction (e) will be included in the limit of bulk-dominated dissipation in section 3.4. Two other assumptions shall also be used later but are not required for the following main derivation:
\begin{itemize}
\item[(f)] The surface displacement falls into the propagative wave regime, i.e. the horizontal sizes of the surface displacement are larger than the viscous length $\ell_\nu = g^{-1/3}{ \nu_\ell}^{2/3}$. 
\item[(g)] The surface displacement $\zeta$  is small compared to the viscous sublayer $\delta_\nu$ of the turbulent boundary layer in the air.
\end{itemize}

Assumption (f) is comfortably satisfied experimentally in the wrinkle regime for usual values of liquid viscosity, $\nu_\ell < 1000 \nu_{\textrm{water}}$. Assumption (g) will be useful in section 4 to model  the pressure and shear applied at the liquid interface by the gas layer.

\subsection{Formulation}

We consider a liquid layer submitted to a normal stress field $N(x,y,t)$ and a shear stress $\bs T(x,y,t) = T_x \bs e_x + T_y \bs e_y$ applied at its upper surface in $z=\zeta(x,y,t)$. The linearized Navier-Stokes equation for the velocity $\bs v = v_x \bs e_x + v_y \bs e_y + v_z \bs e_z$ in the liquid reads
\bb
\partial_t\boldsymbol v &=&- \frac{1}{\rho_\ell} \boldsymbol \nabla p_\ell + \mathbf g + \nu_{\ell}  \Delta \boldsymbol v ,
\label{NS}
\ee
where $p_\ell$ is the pressure in the liquid phase. We have $\boldsymbol \nabla \cdot \boldsymbol v = 0$ and the boundary condition for an infinite liquid depth reads
\bb
 \lim_{z\rightarrow -\infty}\bs v = \bs 0 .
\ee
At the interface in $z=\zeta$, the continuity condition of the normal stress can be approximated by the pressure in $z=0$ using $p_\ell(x,y,z=\zeta) = p_0 - \rho_\ell g \zeta$, where $p_0 = p_\ell(x,y,z=0)$. The continuity condition for the normal stress in $z=0$ yields
\bb
p_0  - \rho_\ell g \zeta - 2{\rho_\ell}\nu_{\ell} (\partial_z v_z)_{z=0}+ {\gamma} \Delta_{(x,y)} \zeta &=& N ,
\label{eq_nstress}
\ee
where $\Delta_{(x,y)}$ is the 2D Laplacian. The continuity condition of the tangential stress writes
\begin{subeqnarray}
{\rho_\ell}\nu_{\ell} \big( \partial_z v_x +\partial_xv_z\big)_{z=0}&=&T_x\\
{\rho_\ell}\nu_{\ell} \big( \partial_z v_y +\partial_yv_z\big)_{z=0}&=&T_y .
\label{eq_tstress}
\end{subeqnarray}

\subsection{Statistically steady and homogeneous regime}

We now look for statistically homogeneous and stationary solutions. We  introduce the space-time Fourier transform $\mathcal F$,
 \begin{subeqnarray}
\hat \zeta({\boldsymbol k},\omega) &=& \mathcal F \{ \zeta({\boldsymbol r},t) \} = \int{\rm d}^2 \boldsymbol r \, {\rm d} t\, \zeta({\boldsymbol r},t) e^{-i(\bs k \cdot \bs r - \omega t)} \\
\zeta({\boldsymbol r},t) &=& \mathcal F^{-1} \{ \hat \zeta({\boldsymbol k},\omega) \} = (2\pi)^{-3} \int{\rm d}^2 \bs k \, {\rm d} \omega\, \hat \zeta({\boldsymbol k},\omega) e^{i(\bs k \cdot \bs r - \omega t)} ,
\label{zetafourier}
\end{subeqnarray}
and similarly for $\hat N({\boldsymbol k},\omega)$ and $\hat {\bs T}({\boldsymbol k},\omega)$, with ${\boldsymbol r} = x {\boldsymbol e}_x + y{\boldsymbol e}_y$ the horizontal position, ${\boldsymbol k} = k_x {\boldsymbol e}_x + k_y {\boldsymbol e_y}$ the horizontal wave vector and $\omega$ the angular frequency ($\bf k$ and $\omega$ are real). From Eq.~(\ref{NS}) we obtain
\bb
\Delta p_\ell &=&0 \label{diffP} \\
\partial_t \bs \Omega- \nu_\ell \Delta \bs \Omega &=&\bs 0 \label{diffOmega} ,
\ee
where we have introduced the vorticity $\bs \Omega =\bs \nabla \times \bs v $. The boundary conditions for $\bs \Omega$ writes
\bb
\bs \Omega(\bs r,z=0,t) &=& {\boldsymbol \Omega}_0({\boldsymbol r},t) \\
\lim_{z\rightarrow - \infty} \bs \Omega(\bs r,z,t) &=& \bf 0,
\ee
where ${\boldsymbol \Omega}_0({\boldsymbol r},t)$ is the vorticity at the surface. Eqs.~(\ref{diffP}) and (\ref{diffOmega}) can be solved in Fourier space,
\bb
 p_\ell({\boldsymbol r},z,t)  &=& (2\pi)^{-3}\int{\rm d}^2 \bs k \, {\rm d} \omega\, \hat {p}_0(\bs k, \omega) e^{i(\bs k \cdot \bs r - \omega t)} e^{kz} \label{pfourier}\\ 
\bs \Omega({\boldsymbol r},z,t) &=& (2\pi)^{-3} \int{\rm d}^2 \bs k \, {\rm d} \omega\, \hat {\bs \Omega}_0(\bs k, \omega) e^{i(\bs k \cdot \bs r - \omega t)} e^{mz} \label{ofourier} ,
\ee
 where
\bb
m^2 = k^2-i\omega/\nu_\ell,
\label{defm2}
\ee
with $k = (k_x^2 + k_y^2)^{1/2}$ and $\{\hat {p}_0, \hat {\bs \Omega}_0\} = \mathcal F \{  {p}_0 ,  {\bs \Omega_0}\}$ being the Fourier transforms of the pressure and vorticity in $z=0$. Eq.~(\ref{pfourier}) shows that an applied pressure patch of typical wavenumber $k$ and frequency $\omega$ penetrates the liquid layer over a depth $k^{-1}$, while Eq.~(\ref{ofourier}) shows that the applied shear stress and hence vorticity penetrate over a depth $|m|^{-1}$. These two penetration depths are similar for very viscous fluids, whereas in the limit of low viscosity vorticity remains confined in a thin boundary layer of thickness $|m|^{-1} \simeq \sqrt{\nu_\ell / \omega}$.

Rewriting Eq.~(\ref{NS}) as $\partial_t\bs v = -\bs \nabla p_\ell/\rho_\ell  -\nu_\ell \bs \nabla \times \bs \Omega$, and using  Eqs.~(\ref{pfourier})-(\ref{ofourier}), we obtain the expression of the velocity
\bb
\bs v ({\bf r},z,t) &=& (2\pi)^{-3}\int{\rm d}^2 \bs k \, {\rm d} \omega\left( \frac{\bs q}{\rho_\ell i \omega} \hat p_0 e^{kz} +  \frac{\nu_\ell}{i \omega}\bs \kappa \times \hat {\bs \Omega}_0 e^{mz}  \right)e^{i(\bs k \cdot \bs r - \omega t)}, \label{velocity}
\ee
where $\bs q = (i k_x, i k_y,k)$ and $\bs \kappa = (i k_x, i k_y,m)$. The functions $\hat {p}_0$ and $\hat {\bs \Omega}_0$ are given by the stress boundary conditions in $z=0$. Combining Eqs.~(\ref{eq_nstress}), (\ref{eq_tstress}) and (\ref{velocity}) yields
\bb
\left ( 1 - \frac{2 k^2 \nu_\ell}{i \omega} \right )\frac{k}{\rho_\ell} \hat p_0- \frac{2 \nu_\ell m k  }{i \omega} \nu_\ell \hat B_z -g' k \hat \zeta &=& \frac{k \hat N}{\rho_\ell} \label{eq_nstressF}\\
\nu_\ell \hat B_z+ 2 \nu_\ell\mathcal{F}\{(\partial_{zz} v_z)_{z=0}\} &=&- \frac{i \bs k \cdot \hat{\bs T}}{\rho_\ell} , \label{eq_tstressF}
\ee
where $g' = g + \gamma k^2/\rho_\ell$ is the modified gravity and $\hat B_z =  (\bs \kappa \times \hat {\bs \Omega}_0(\bs k, \omega)) \cdot \bs e_z$ satisfies $\Delta v_z = - B_z$. The component $B_z$ is  associated to the viscous dissipation of the vertical component $v_z$ of the velocity field. Using the kinematic condition $\partial_t \zeta = (v_z)_{z=\zeta}$ in the small perturbation limit, we obtain the relation between $\hat \zeta, \hat p_0$ and $\hat B_z$,
\bb
\omega^2 \hat \zeta = \frac{k}{\rho_\ell} \hat p_0 +  \nu_\ell \hat B_z \label{eq_zpB} ,
\ee
where $\hat p_0$ and $\hat B_z$ can be expressed using Eqs~(\ref{eq_nstressF}) and (\ref{eq_tstressF}). From Eqs.~(\ref{eq_nstressF}), (\ref{eq_tstressF}) and (\ref{eq_zpB}) we obtain by algebric manipulations (see details in Appendix A) the following expression
\bb
\left (\omega^2 - g' k + 4 i \nu_\ell \omega k^2 + 4 \nu_\ell^2 k^3 (m-k) \right) \hat \zeta &=& \frac{k \hat N}{\rho_\ell} + \frac{m-k}{m+k}\frac{i \bs k \cdot \hat{\bs T}}{\rho_\ell} .
\label{eq_perrard}
\ee
We obtain a damped wave equation forced by the normal and tangential stresses applied at the interface. It describes the response of a viscous liquid forced by arbitrary normal and tangential stress fields under the assumptions (a) to (e), in statistically steady and homogeneous configurations. Setting $N$=0 and ${\bf T} = {\bf 0}$ yields the usual dispersion relation for gravity-capillary waves with viscosity~\citep{Lamb_1995}. In a very viscous fluid ($m \simeq k$), the effect of the shear stress on the flow vanishes, whereas for a fluid of low viscosity ($|m| \gg k$), the effects of pressure and shear stress are comparable. Note here the specificity of the limit $\nu_\ell \rightarrow 0$: although $(m-k)/(m+k) \rightarrow 1$, only pressure can generate waves since the inviscid limit implies $\bf T = \bs 0$ by tangential stress continuity.

Eq.~(\ref{eq_perrard}) can be simplified using the small viscosity approximation (assumption (f)). The boundary layer thickness associated to the vertical diffusion of a shear stress patch applied during a time $\omega^{-1}$ is $\delta_\ell = \sqrt{\nu_\ell/\omega}$. For a gravity wave of frequency $\omega = \sqrt{g k}$ we have $\delta_\ell = \ell_\nu^{3/4} k^{-1/4}$, where $\ell_\nu=(\nu_{\ell}^2/g)^{1/3}$ is the viscous length introduced in section 2.1. In practice, $\delta_\ell$ lies in the range 0.2 - 7 mm for a typical wrinkle wavelength $2\pi/k \approx 100$~mm. The thin boundary layer approximation (f) is therefore fulfilled. In this limit, we have $|m| \gg k$, and only the first order in $\nu_\ell$ contributes in Eq.~(\ref{eq_perrard}), yielding
\bb
\hat \zeta ({\bs k},\omega) =\frac{1}{\rho_\ell} \frac{k \hat N + i \bs k \cdot \hat{\bs T}}{\omega^2 - g' k + 4 i \nu_\ell \omega k^2} .
\label{eq_psimp}
\ee
Equation~(\ref{eq_psimp}) is the corner stone of this paper: it relates the Fourier component of the displacement field $\hat \zeta ({\bs k},\omega) $ to the Fourier components of the applied normal and tangential stresses $\hat N$ and $\hat {\bs T}$. \textcolor{black}{The surface response in the physical space is finally obtained by applying the inverse Fourier transform (\ref{zetafourier}b) to Eq.~(\ref{eq_psimp}),
\bb
\zeta ({\bs r},t) = \frac{1}{(2\pi)^3} \frac{1}{\rho_\ell} \int{\rm d}^2 \bs k \, {\rm d} \omega\  \frac{k \hat N + i \bs k \cdot \hat{\bs T}}{\omega^2 - g' k + 4 i \nu_\ell \omega k^2} .
\label{eq_psimpreal}
\ee 
In the specific case of $\hat {\bs T} ({\bs k},\omega)= {\bf 0}$ and $\hat N ({\bs k},\omega)$ in the form $\delta(\omega - U_c k_x) \hat {\cal N}({\bs k})$, with $U_c$ the convection velocity of a rigid pressure source of Fourier transform $\hat {\cal N}({\bs k})$, this equation reduces to the classical Havelock integral used to describe the far-field wake of a ship~\citep{havelock1919,Raphael_1996}.}

\subsection{Interpretation}

\textcolor{black}{Equation~(\ref{eq_psimp}) can be analysed from the point of view of the linear response theory. It can be written in the form
\bb
\hat \zeta(\bs k,\omega) = \frac{\hat S(\bs k,\omega)}{\hat D(\bs k,\omega)},
\label{eq_pratio}
\ee
with the source term defined as
\bb
\hat S({\bs k},\omega) = \frac{k \hat {N} + i \bs k \cdot \boldsymbol{ \hat {T}}}{\rho_\ell},
\ee 
and the (inverse) spectral convolution kernel $\hat D$ as
\bb
\hat D(\bs k,\omega) = \omega^2 - g' k~ \textrm{tanh}(k h) + 4 i \nu_\ell \omega k^2.
\label{eqD}
\ee
The real part of $\hat D(\bs k,\omega)=0$ corresponds to the inviscid dispersion relation for gravity-capillary waves, generalized here to arbitrary depth $h$. This generalisation is valid in the limit of bulk-dominated dissipation for small viscosity (assumption (g)).}
 
\begin{figure}
\centering
\includegraphics[width=0.5 \columnwidth]{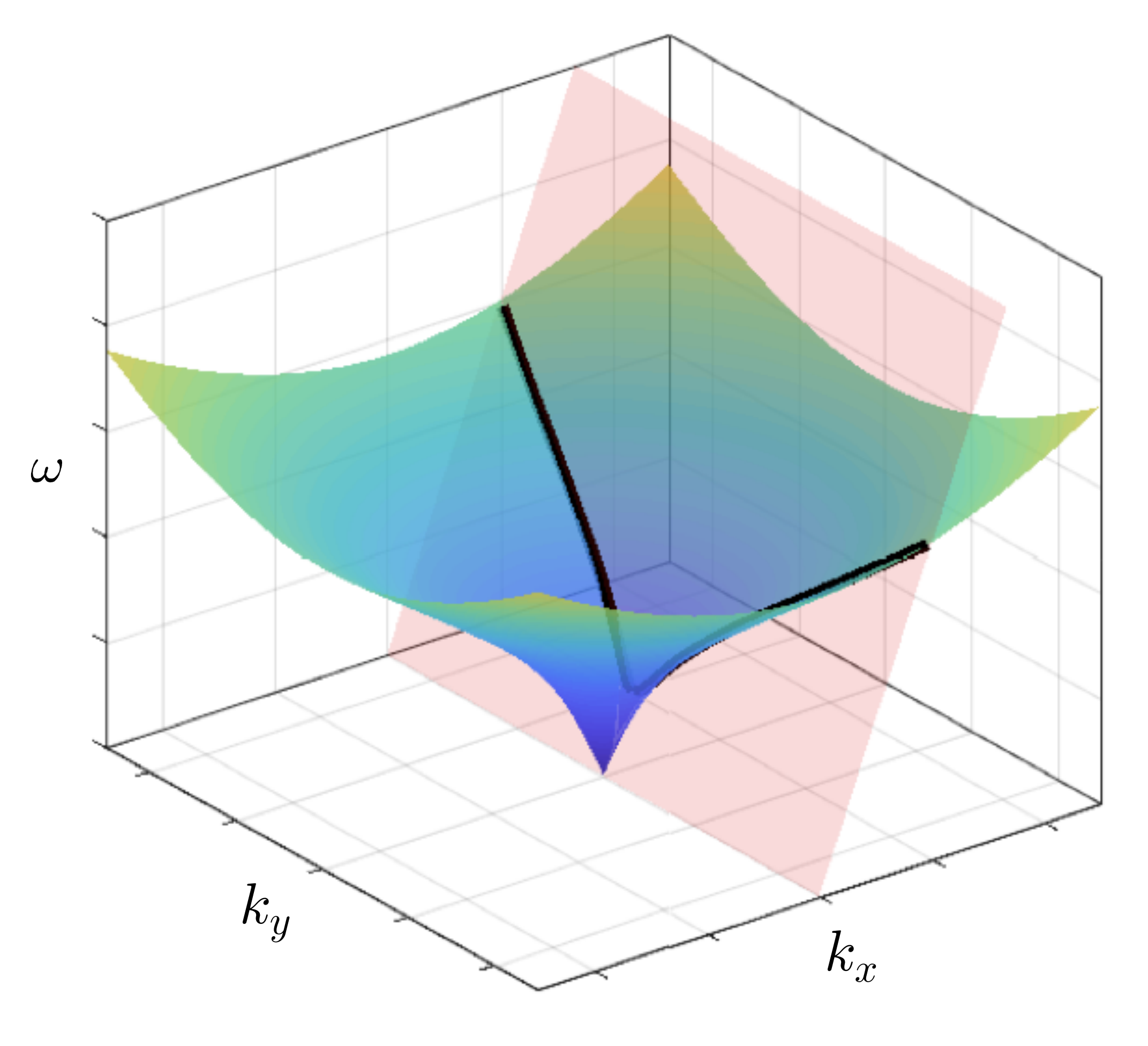}
\caption{Representation in the Fourier space of the  dispersion relation $\omega = \sqrt{g' k \tanh (kh)}$ of gravity-capillary waves (blue-green surface) and the forcing (pink plane). The forcing here corresponds to a source traveling at constant velocity in the $x$ direction,  $\omega = U_\mathrm{ c} k_x$. The intersection between the two surfaces (black line) is where the maximum wave amplitude is expected.}
\label{fig1}
\end{figure}

\textcolor{black}{In the Fourier space $(k_x,k_y,\omega)$, the inviscid dispersion relation forms a surface with rotational invariance around the $\omega$ axis, represented  by the blue green surface in figure~\ref{fig1}. The Fourier modes $(\bs k,\omega)$ in the immediate vicinity of the surface $\hat D = 0$ correspond to propagative waves. Although the finite depth and the capillary terms are not expected to play a key role in the wrinkle generation (see section 2),  we briefly recall here the main properties of the dispersion relation in the general case.  Three propagation regimes may be defined: $k h \ll 1$ corresponds to the shallow water regime, in which surface waves are non dispersive (finite slope at the origin in figure~\ref{fig1}, given by the phase velocity $c = \omega/k = \sqrt{gh}$). The two other propagation regimes, lying in the deep water domain $k h \gg 1$, have a phase velocity $c^2 = g (1/k + k \ell_\mathrm{c}^2)$, with $\ell_c$ the capillary length. For $k \ell_c < 1$, gravity effects dominate over capillary forces, while for $k \ell_c > 1$ capillary forces take over, with a change of curvature in $\omega$ at $k \ell_c = 1$. At this inflection point the phase velocity is minimum, $c_{\mathrm{min}} = (4 g \gamma/\rho_\ell)^{1/4}$.  For the interface between air and water or the viscous aqueous solutions considered here, this minimum phase velocity is $c_{\mathrm{min}} \simeq 22$~cm/s.}

\textcolor{black}{In view of Eq.~(\ref{eq_pratio}), a significant surface response is expected where $\hat D$ is small and $\hat S$ is large. More specifically, Eq.~(\ref{eq_pratio}) predicts an excitation of the Fourier modes $(\bs k,\omega)$ supplied by the source $\hat S$ that fall in the vicinity of the dispersion relation. In the simple case of pressure or stress patches rigidly travelling at a constant velocity $U_\mathrm{c}$ in the $x$ direction, $\hat S$ is non-zero for $\omega = k_x U_\mathrm{c}$, represented as a tilted plane in pink in figure~\ref{fig1}. The intersection  between this tilted plane and the surface $\hat D = 0$ (shown as a black curve), which is defined provided that $U_c \geq c_{\mathrm{min}}$, is naturally where wave amplification is expected. For such idealised 'rigid' forcing, the resulting wave pattern is a collection of wakes, stationary in the frame of the source (i.e., with a convection velocity given by $U_c$), analogous to far-field wakes behind ships.}

\textcolor{black}{In the case of time-varying forcing, relevant to the problem of a turbulent wind blowing over the liquid surface, the forcing spectrum is now a continuum of Fourier modes centred around $\omega \simeq U_c k_x$, in a subspace of thickness along $\omega$ given by the inverse correlation time of the fluctuations. Loosely speaking, the black line in figure~\ref{fig1} now has a finite thickness, allowing in principle for wave excitation even if $U_c < c_{\mathrm{min}}$. We expect the wrinkles to be the unstationnary wakes generated by this turbulent forcing. For a more quantitative description, the spatio-temporal properties of the forcing $\hat S$ must be specified, to which section~\ref{sec:dns} is devoted.}

\textcolor{black}{It is worth noting that the physical picture given here for the wrinkle regime corresponds only to the {\it linear} response of the free surface, and ignores  energy transfers between modes of finite amplitude and retroaction on the turbulent forcing. The {\it regular waves}, defined as the subset of Fourier modes satisfying $k_y=0$ (i.e., propagating in the streamwise direction), which are found experimentally at large wind velocity, probably escape from this linear description. Such regular waves first appear with a phase velocity of the order of $c_{\mathrm{min}}$, which is much smaller than the forcing velocity $U_c$, suggesting that they originate from a retroaction of the liquid surface on the turbulent forcing. We shall return to this point in section~\ref{sec:orw}.}

\subsection{Non dimensional form}

We express now the surface response (\ref{eq_psimp}) in the form~(\ref{eq:f3}) inferred from the dimensional analysis. Using the boundary layer thickness $\delta$ as the characteristic length and $\delta / u^*$ as the characteristic time, we introduce the dimensionless source term $\hat S^\dagger$ as
\bb
{\hat S}^\dagger = \frac{\rho_\ell}{\rho_a}\frac{\hat S}{{u^*} \delta^2}.
\label{eq_Sd}
\ee
The mean surface displacement $\overline{\zeta^2}$ can be related to its representation in Fourier space using Parseval's theorem,
\textcolor{black}{
\bb
 \int {\textrm d}^2{\boldsymbol r}\textrm dt \, \zeta^2 = \frac{1}{(2\pi)^3} \int {\textrm d}^2{\boldsymbol k} \textrm d\omega \, |\hat \zeta|^2.
\label{eq_parseval}
\ee
}
Replacing $\hat \zeta$ in Eq.~(\ref{eq_parseval}) by its expression from Eq.~(\ref{eq_pratio}) and Eq.~(\ref{eq_Sd}) yields
\textcolor{black}{
\bb
\frac{\overline{\zeta^2}}{\delta^2} = \left (\frac{\rho_\mathrm{a}}{\rho_\ell} \right )^2 \frac{{u^*}^3}{\delta} \frac{1}{(2\pi)^3}  \int {\textrm d}^2{\boldsymbol k}\textrm d\omega ~ \frac{|{\hat S^\dagger}|^2}{(\omega^2-\omega_r^2)^2 + \omega_\nu^2 \omega^2} ,
\label{eq_pvsol}
\ee
}
where $\omega_r = \sqrt{g' k ~ \textrm{tanh}(k h)}$ is the surface wave frequency and $\omega_\nu = 4 \nu_\ell k^2$ is the dissipation rate. Equation~(\ref{eq_pvsol}) needs further specifications to be written in a dimensionless form. A choice arises for the dimensionless frequency, which can be constructed either from the characteristic time of the source, $\delta/u^*$, or from the characteristic time of the surface response, ${\omega_\nu}^{-1}$. We chose the dimensionless frequencies $\tilde \omega = \omega/\omega_\nu$, $\tilde \omega_r = \omega_r/\omega_\nu$ and the dimensionless wavenumber $\tilde k =  k \delta$, yielding
\bb
\frac{\overline{\zeta^2}}{\delta^2} &=& \left (\frac{\rho_\mathrm{a}}{\rho_\ell} \right )^2 \frac{{u^*}^3}{4 g \nu_\ell}  \frac{1}{(2\pi)^3} \nonumber \\
&\times&  \int  \textrm d^2{\boldsymbol {\tilde k}}\textrm d\tilde \omega ~ \frac{1}{\tilde k^3 (1+ \mathrm{Bo}^{-2} \tilde k^2 ) \tanh (\tilde k h/\delta )} ~~ \frac{|{\hat S^\dagger}|^2}{(\tilde \omega-\tilde \omega_r)^2 (\tilde \omega / \tilde \omega_r +1)^2 + (\tilde \omega / \tilde \omega_r)^2} . \quad
\label{eq_pvol_dless}
\ee
This equation confirms the role played by the dimensionless combination ${u^*}^3/(g \nu_\ell)$ introduced in section 2, and provides an analytical expression for the dimensionless function $f_3$ in Eq.~(\ref{eq:f3}). However, the dependency in Reynolds number is hidden in the source term $\hat S^\dagger$. Hence, a quantitative description of the spectral source originating from the turbulent boundary layer in the air phase is now required.

\section{Properties of the turbulent forcing from DNS}
\label{sec:dns}

We compute the source term $\hat S$ using a set of time-resolved pressure and shear stress fields evaluated at $z=0$, taken from three-dimensional DNS of a developed turbulent flow in a non deformable channel with \textcolor{black}{no-slip condition at the bottom and top boundaries, and periodic boundary conditions along the streamwise and spanwise directions}. In order to provide comparison with the experiments of \cite{Paquier_2015,Paquier_2016}, which were performed in a developing boundary layer flow, we assume here that the spatio-temporal statistics of turbulence in a developing boundary layer of thickness $\delta$ at a given Re$_\delta= \delta u^* / \nu_a$ are equivalent to that of a channel flow of half-height $H$ at the same value of Re$_\tau = H u^* /\nu_a$. Previous works have shown that this assumption is reasonably well satisfied for the flow close to the wall, $z < 0.6 \, \delta$~\citep{Jimenez_2008,Jimenez_2010}. In the following, we identify $H$ to $\delta$ and we use for simplicity the same notation Re$_\delta$ for the DNS and the experiments.

\subsection{Boundary conditions}

We first examine to what extent a canonical turbulent flow over a smooth and rigid wall with no slip boundary condition can adequately model the turbulence over a free surface.   We base our analysis of the air flow on three assumptions:
\begin{itemize}
\item The interface is slightly deformable, $\partial_t \zeta|_{z=0^+} = v_z|_{z=0^+}  \ll v_{x,y}|_{z=0^+}$.
\item The interface is smooth, $\zeta \ll \delta_\nu$.
\item The surface drift velocity in the liquid is negligible compared to the velocity of the turbulent structures, $v_x|_{z=0^-} \ll U_a$.
\end{itemize}
These three assumptions are corollary of the assumptions (b), (d) and (g) discussed in section 3.1. The rigid wall approximation is equivalent to the linear approximation (b). The smooth wall approximation is motivated by the typical wrinkle amplitude $\zeta_{\textrm{rms}}$, at least ten times smaller than the viscous sublayer $\delta_\nu$ below the wind wave threshold~\citep{Paquier_2016}. It is therefore a consequence of assumption (g). Finally, the no-slip boundary condition derives from assumption (d).

Under these assumptions, we model the turbulence by a turbulent channel flow over a smooth and rigid wall with no slip boundary condition in $z=0$ and $z=2H$ and periodic boundary conditions along $x$ and $y$. The DNS configuration is sketched in figure~\ref{setup}, with $L_x, L_y$ the streamwise and the spanwise lengths. The flow is driven by a mean streamwise pressure gradient $-P_a/L_x$. By conservation of the streamwise momentum, the mean tangential stress at the boundaries is $\tau_a = H P_a /L_x$, which defines the friction velocity $u^* = \sqrt{\tau_a/\rho_a}$. We decompose the instantaneous pressure at the wall as the sum of a stationary pressure drop $P(x) = P_a (1-x/L_x)$ and turbulent pressure fluctuations $p(x,y,t)$ of zero mean. Similarly, the wall shear stress is the sum of a stationary component $\tau_a {\boldsymbol e_x}$ and  fluctuations ${\boldsymbol \sigma}(x,y,t) = \sigma_x {\boldsymbol e_x} + \sigma_y {\boldsymbol e_y}$ of zero mean. As argued before, the steady contributions are responsible for a mean flow generation in the liquid, which is not considered here. In the following, we focus on the fluctuating contributions $(p, {\boldsymbol \sigma})$.

To compute the source term $\hat S = (k \hat N + i \bs k \cdot \hat{\bs T}) / \rho_\ell$ from the $(p, {\boldsymbol \sigma})$, we need to specify the normal and tangential stresses,
\bb
N &=&  p + 2 \rho_a \nu_a (\partial_z v_z)_{|z=0^+} \\
T_x &=&  \sigma_x + \rho_a \nu_a  (\partial_x v_z)_{|z=0^+} \\
T_y &=&  \sigma_y +  \rho_{a} \nu_{a}  (\partial_y v_z)_{|z=0^+}.
\ee
The relative importance of the viscous contributions depends on the surface deformation and the magnitude of the turbulent air flow. For a surface displacement $\zeta$ of characteristic wavenumber $k$ and convection speed $U_c$, spatial gradients of $v_z$ scale as $\partial_{x,y,z} v_z \approx \zeta U_c k^2$. For a turbulent boundary layer of friction velocity $u^*$, the vertical gradient of horizontal velocity scales as $\partial_z v_{x,y} \approx u^*/\delta_\nu$. The ratio $\partial_{x,y} v_z/\partial_z v_{x,y}$ is thus given by $\zeta \delta_\nu u^* k^2/U_c$, where $U_c \approx 12 u^*$. If we consider a wave of wavelength $\Lambda$, using assumption (b), $\zeta \ll \Lambda$, and assumption (g), $\delta_\nu \ll \Lambda$, we have in practice $\partial_{x,y,z} v_z \ll \partial_z v_{x,y}$ in the wrinkle regime. The expressions of $N$ and $\bs T$ thus reduce to
\bb
N &=& p \\
\bs T &=& \bs \sigma,
\ee 
where $p$ is the air pressure and $\bs \sigma$ is the tangential wall stress in the limit of a non deformable wall. In the following, we  use the conventional wall-unit notation +,
\bb
\{p,\bs \sigma\} &=& \rho_{a} {u^*}^2 ~ \{p^+, \bs \sigma^+\},
\ee
so the dimensionless source term ${\hat S}^\dagger$ defined in Eq.~(\ref{eq_Sd}) reads
\bb
{\hat S}^\dagger ({\bs k},\omega) = \tilde k {\hat p}^+  + i \tilde{\bs k} \cdot  \hat{\bs \sigma}^+,
\label{eqS}
\ee
with $\tilde{\bs k} =  \bs k \delta$.

\subsection{DNS Configuration}

\begin{table}
\begin{center}
 \begin{tabular}{lc c c c c c || c} 
Re$_\delta$ & $\Delta x^+$ & $\Delta y^+$ & $\Delta z_{min}^+$ & $\Delta z_{max}^+$ & $\Delta t^+$  & $\textcolor{black}{T_{max} u^* / \delta}$ & $U_a$ (m~s$^{-1}$)\\ [0.5ex] 
100      &    10.1      &    5.7      &     0.06            &     3.4            &   0.63       &   12.5     & 1 \\ 
180      &     9.1      &    5.3      &     0.02            &     3.0            &   0.64       &   14.1     & 1.8 \\
 250      &    12.1      &    6.8      &     0.03            &     4.0            &   0.61       &   10.1     & 2.5 \\
  360      &    13.1      &    6.5      &     0.04            &     5.8            &   3.80       &   21.8     & 3.6 \\
  550      &    13.4      &    7.5      &     0.04            &     6.7            &   0.45       &    6.7     & 5.5 \\ 
  \label{table2}
\end{tabular}
\end{center}
   \caption{Details of the DNS turbulent channel flow. $\Delta x^+$ and
     $\Delta y^+$ are the spatial resolutions in terms of Fourier modes
     before dealiasing. $\Delta z_{min}^+$ and $\Delta z_{max}^+$ are the
     finest and coarsest spatial resolutions in the wall-normal
     direction in dimensionless units. $\Delta t^+$ is the temporal separation between stored
     flow fields and $T_{max}$ is the total time simulated. $U_a$ is the
     corresponding wind velocity for a boundary layer
     thickness $\delta \simeq 30$~mm and a kinematic
     viscosity of air $\nu_a = 15\times
     10^{-6}$~m$^2$~s$^{-1}$ (see section 2.2).}
\end{table}

\begin{figure}
\centering
\includegraphics[width=0.9\columnwidth]{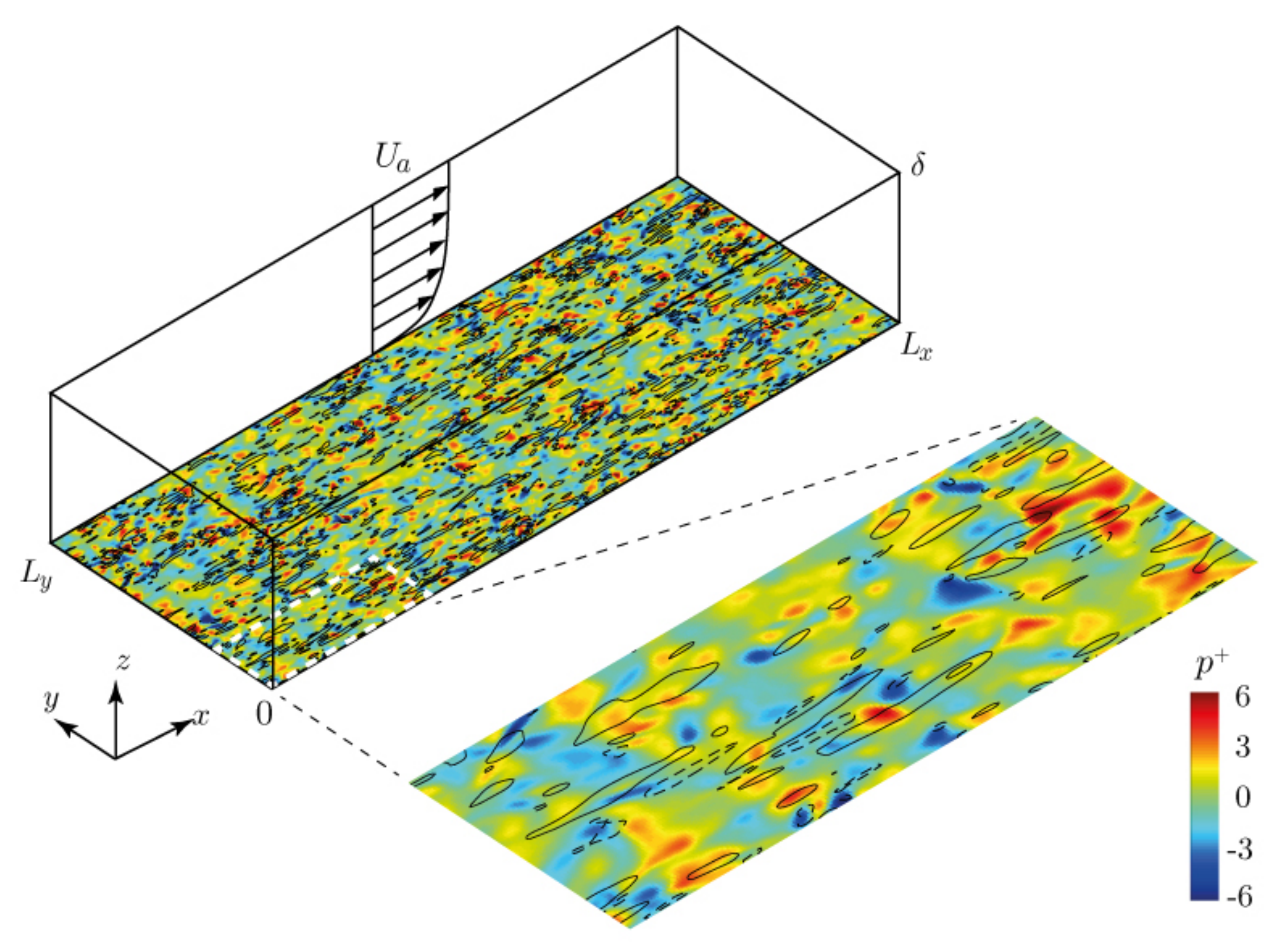}
\caption{Sketch of the numerical setup. The pressure and the shear stress on the wall plane $z=0$ taken from Direct Numerical Simulation of a turbulent channel flow are applied to the surface of a viscous liquid. The DNS is performed on a domain $(L_x, L_y, L_z) = (8\pi, 3\pi, 2) \delta$. The snapshot illustrates the turbulent field in $z=0$ for the case Re$_\delta = 250$. Pressure $p^+ = p / (\rho_a u^{*2})$ is shown in color, and longitudinal shear stress $\sigma_x^+$ as contour lines (lines are separated by increments $\sigma_x^+ = 1$, positive for full lines and negative for dashed lines). A magnification of the snapshot by a factor 4 in each direction is also presented.}
\label{setup}
\end{figure}

\textcolor{black}{The numerical set up configuration is sketched in figure~\ref{setup}, with the wall pressure and wall shear stress taken for Re$_\delta$=250. The parameters for each DNS run are summarised in table~2. The turbulent Reynolds number $\textrm{Re}_\delta$ ranges from 100 to 550, which corresponds in the experiments of \citet{Paquier_2015} to wind speeds ranging from 1 to 5.5~m~s$^{-1}$. This correspondence is obtained by equating the DNS and experimental values of $\textrm{Re}_{\delta}$, with $\nu_a = 15 \times 10^{-6}$~m~s$^{-2}$ for the kinematic viscosity of air, and $\delta \simeq 30$~mm for the local boundary-layer thickness at the $x$-location where measurements are carried out (see section 2.2).}

The incompressible flow is integrated in the form of evolution equations for the wall-normal vorticity and for the Laplacian of the wall-normal velocity, as in \citet{kim:moi:mos:87}, and the spatial discretisation is desaliased Fourier series in the two wall-parallel directions and Chebychev polynomials in $z$. Time stepping is the third-order semi-implicit Runge-Kutta method from \citet{Moser_1999}.

\textcolor{black}{The computational box is $L_x \times L_y \times L_z = (8\pi, 3\pi,2)\delta$ with periodic boundary conditions along $x$ and $y$ directions. These spatial dimensions are larger than the minimum channel size $(2 \pi, \pi, 2)\delta$ often used in numerical simulation~\citep{Jimenez_2013}.  Preliminary tests showed that this large domain size is necessary to ensure the correct convergence of the Fourier integral (\ref{eq_psimpreal}), which is dominated by the contributions at small $\bs k$. }

\textcolor{black}{The periodic boundary condition in time, implicitly assumed in our spectral formulation, is naturally not satisfied in the DNS data. However, spurious temporal correlations are limited by the large computational domain: the correlation time of the pressure fluctuations, of order of $20 \mathrm{Re}_\delta^{-1} \delta/u^*$, is comfortably smaller than the transit time over the computational domain, of order of $L_x/U_c \simeq 2 \delta/u^*$, where $U_c$ is the convection velocity (see section~\ref{sec:stats}). The total integration time $T_{max}$ is chosen at least $10 \delta/u^*$ (except for the largest Re$_\delta$), to correctly resolve the lowest frequencies $\omega$ in the wave dynamics. The time step, $\Delta t^+$, is sufficiently small to resolve the fastest waves (the case Re$_\delta$ = 360 has a coarser time step in order to collect statistics for a longer time period).}

\subsection{Pressure and shear stress statistics at the wall}
\label{sec:stats}

\begin{figure}
\centering \includegraphics[width=0.55\columnwidth]{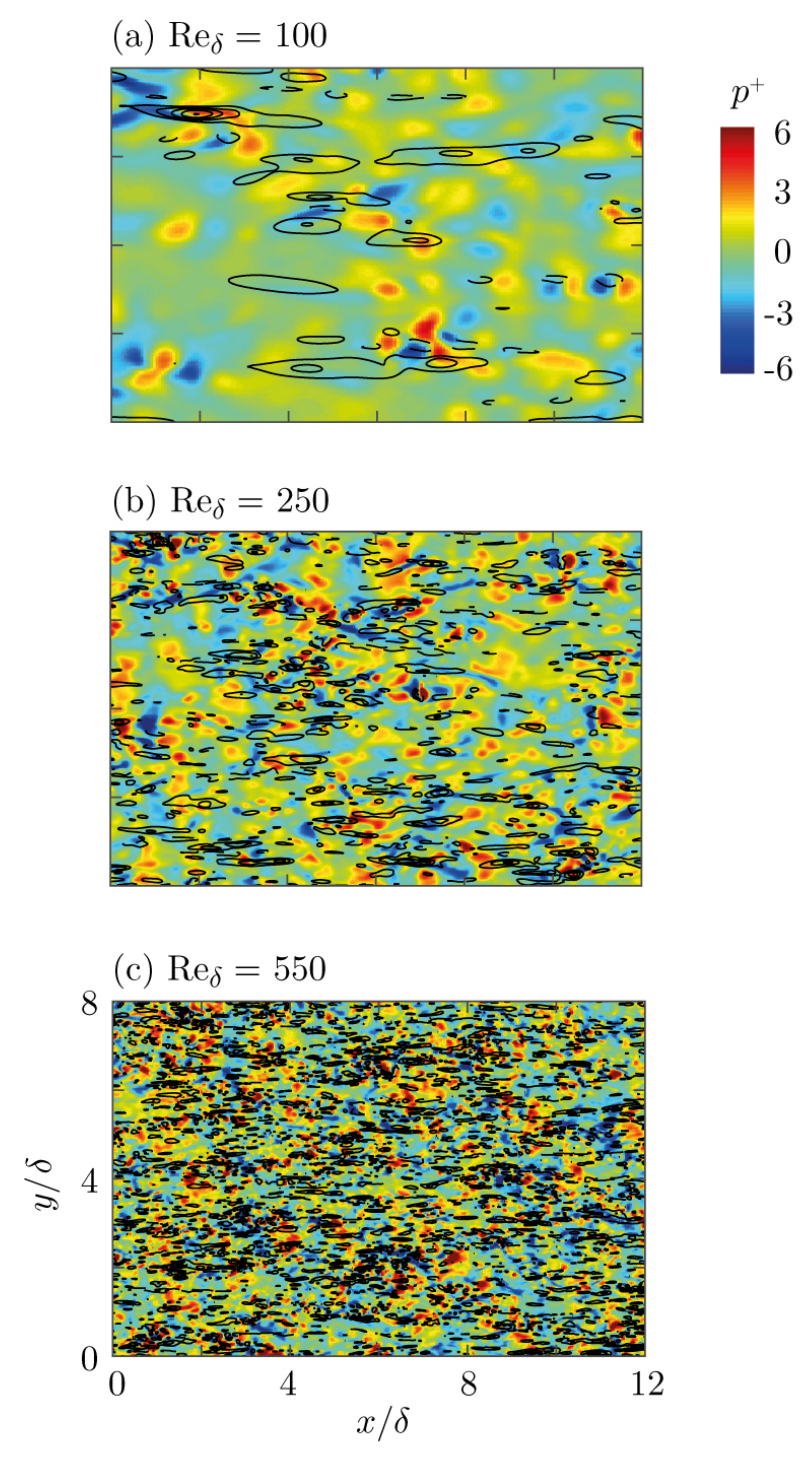}
\caption{Snapshots of the pressure and shear stress fields at the surface, at three values of Re$_\delta$. Only a subdomain $[0, 12 \delta] \times [0, 8 \delta]$ is shown. The pressure $p^+ = p  / (\rho_a u^{*2}$) is shown in colour, and the longitudinal shear stress $\sigma_x^+$ as contour lines, such that $| \sigma_x^+ | = 0.5 i$ with integer $i$ (positive for full lines and negative for dashed lines).}
\label{fig2}
\end{figure}

Figure~\ref{fig2} shows snapshots of the instantaneous wall pressure $p^+$ and streamwise wall shear stress $\sigma_{x}^+$ for Reynolds number Re$_\delta = 100, 250$ and 550. Increasing the Reynolds number naturally decreases the size of the structures. These snapshots show that the shear stress patterns tend to be elongated in the streamwise direction, whereas the pressure patterns are nearly isotropic in the $(x,y)$ plane. These elongated shear stress patterns are a classic signature of the streamwise streaks in the near-wall region of the boundary layer, whereas the nearly isotropic pressure patterns are related to the imprint created at the wall by the cores of the vortices~\citep{Jimenez_2013}.

\begin{figure}
\centering \includegraphics[width=0.95\columnwidth]{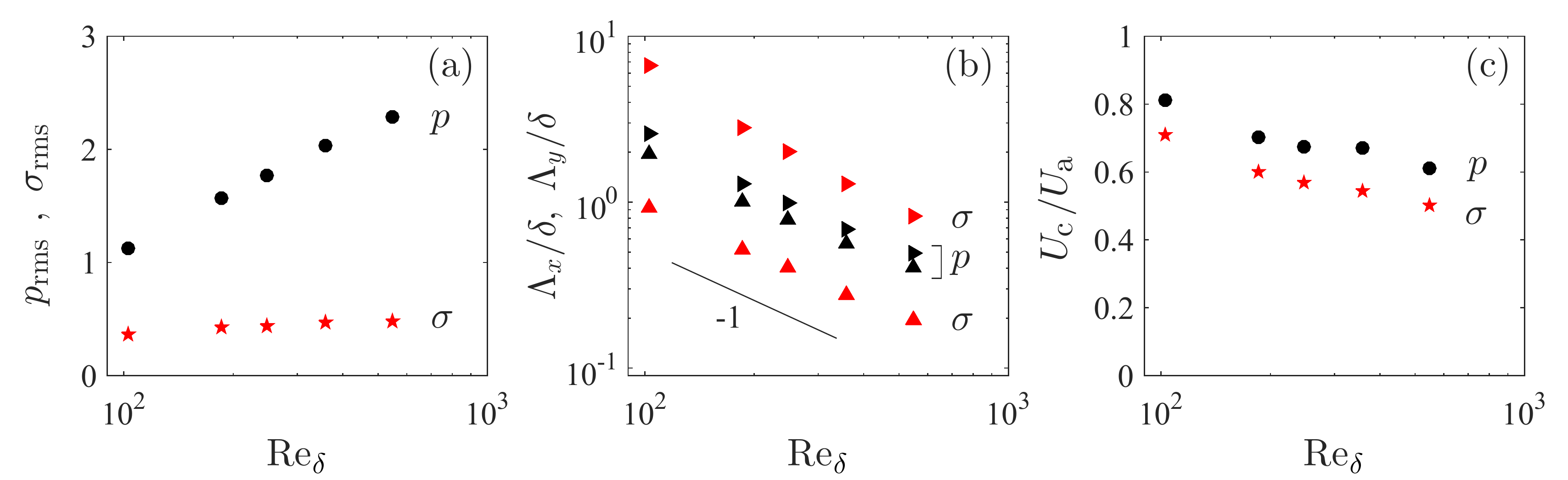}
\caption{Statistics of pressure and shear stress fluctuations as a function of $\textrm{Re}_\delta$.  (a) Root-mean-squared pressure $p_{\textrm{rms}}^+$ ($\bullet$) and stress $\sigma_{\textrm{rms}}^+ = ({{\sigma_x^+}_{\textrm{rms}}}^2 + {{\sigma_y^+}_{\textrm{rms}}}^2)^{1/2}$ ($\textcolor{red}{\star}$). (b) Characteristic streamwise and spanwise lengths $\Lambda_x/\delta$ (\textcolor{black}{$\blacktriangleright$}, \textcolor{red}{$\blacktriangleright$}) and $\Lambda_y/\delta$ ({\small \textcolor{black}{$\blacktriangle$}}, {\small \textcolor{red}{$\blacktriangle$}}) of pressure and shear stress, computed from the spectral barycenters~(\ref{defK}); for $\sigma$, the values of $\Lambda$ are averaged over the two components $\sigma_x$ and $\sigma_y$. Structure sizes decrease as $\Lambda \propto \textrm{Re}_\delta^{-1}$ (black solid line). (c) Mean convection velocity $U_c/U_a$ for pressure ($\bullet$) and shear stress ($\textcolor{red}{\star}$), computed using Eq.~(\ref{defUc}).}
\label{fig3}
\end{figure}

The intensities and sizes of the pressure and shear stress fluctuations are quantified in figure~\ref{fig3} as a function of the Reynolds number. The pressure r.m.s. (figure~\ref{fig3}(a)) is typically 2 to 5 times larger than the shear stress r.m.s. and increases weakly with Re$_\delta$, while the shear stress r.m.s. remains nearly constant over the range of Re$_\delta$ considered. We can therefore anticipate that the surface response will be dominated by the pressure forcing.

To compute the characteristic dimensions of the pressure and shear stress structures, we define for any field $f({\bf r},t)$ the spectral barycenter $\bs K$,
\textcolor{black}{
\bb
\bs K = K_x {\bf e_x} + K_y {\bf e_y} &=& \frac{\int_{\mathcal{D}} {\rm d}^2 \bs k {\rm d}\omega ~ \bs k |\hat f|^2}{\int_{\mathcal{D}} {\rm d}^2 \bs k {\rm d}\omega ~  |\hat f|^2},
\label{defK}
\ee
}
where $\hat f(\bs k, \omega)$ is the Fourier transform of $f$, and ${\mathcal{D}} = \{(k_x,k_y, \omega) | k_x > 0, ~k_y > 0\}$ is the domain of integration.  The mean structure size in the streamwise and spanwise directions, defined as $\Lambda_x = 2\pi/K_x$ and $\Lambda_y=2\pi/K_y$, are plotted in figure~\ref{fig3}(b) for the pressure and the shear stress. $\Lambda_x$ and $\Lambda_y$ both decrease as ${\textrm{Re}_\delta}^{-1}$, indicating that they scale as the (inner) viscous sublayer thickness $\delta_\nu$. \textcolor{black}{The sizes normalized by $\delta_\nu$ are $\Lambda_x^+ \simeq \Lambda_y^+ \simeq 250$ for the pressure patches, and $\Lambda_x^+ \simeq 700$, $\Lambda_y^+ \simeq 100$ for the shear stress patches.}

Similarly, we can define the frequency barycenter $\Omega$ of a field $f$ as
\textcolor{black}{
\bb
\Omega = \frac{\int_{\mathcal{D}} {\rm d}^2 \bs k {\rm d}\omega \, \omega |\hat f|^2}{\int_{\mathcal{D}}{\rm d}^2 \bs k {\rm d}\omega \, |\hat f|^2}.
\label{defO}
\ee
}
We  finally define the convection velocity $U_\mathrm{c}$ as
\bb
U_\mathrm{c} = \Omega/K_x.
\label{defUc}
\ee

\textcolor{black}{The convection velocity $U_c$ for the pressure and the shear stress, plotted in figure~\ref{fig3}(c), lies in the range $[0.5, 0.8] \, U_a$. It slightly decreases with Re$_\delta$,  down to $U_c \approx 0.6 U_a$ for the pressure and  $0.5 U_a$ for the shear stress. These values correspond to the mean velocity at the wall-normal location $z^+ \simeq 30$ where the turbulent fluctuations are maximum~\citep{Kim1989}. Note however that this convection velocity is an average over Fourier components traveling at different velocities: the largest structures propagate at $U_c \approx 0.8 U_a$ while the small scale structures propagate at a slightly lower value, $U_c \approx 0.6 U_a$, as observed experimentally \citep{Willmarth_1962,Corcos_1963} and numerically \citep{Choi_1990}.}

\section{Integrated model of the wrinkle regime}

We now combine the analytical results for the surface response (section 3) with the DNS of the turbulent boundary layer (section 4) to determine the statistical properties of the wrinkles. We first compute in section 5.1 the spatio-temporal fields of \textit{synthetic wrinkles} from direct integration of Eq.~(\ref{eq_psimpreal}) using three-dimensional discrete Fourier transform, and compare them to experimental data. Although a good qualitative agreement is obtained, this direct method suffers from discretisation effects at small wavenumber. To circumvent this difficulty,  we analyse the surface response in the spectral space in section 5.2, and we introduce in section 5.3 a semi-analytical method to evaluate the three-dimensional integral from its dominant contribution in the vicinity of the two-dimensional resonant manifold. This method gives more insight into the physics of the wrinkles and their scaling properties.

\subsection{Surface displacement computation}

We first provide here a direct computation of time series of \textit{synthetic wrinkles} fields from direct integration of Eq.~(\ref{eq_psimpreal}).  From the space-time Fourier transform of the wall pressure $\hat p^+ ({\boldsymbol k},\omega)$ and  wall shear stress $\hat {\bs \sigma}^+ ({\boldsymbol k},\omega)$ extracted from the DNS runs, we compute the source term ${\hat S}^\dagger ({\boldsymbol k},\omega)$ from Eq.~(\ref{eqS}) on a discrete three-dimensional Cartesian grid $(k_x, k_y,\omega)$. Since the surface response occurs mainly at low wavenumber and low frequency, we perform a spectral decimation: we retain only the modes $k_i \delta < 20$ and $\omega \delta/ U_a < 25$.

\begin{figure}
\centering
\includegraphics[width=0.9\columnwidth]{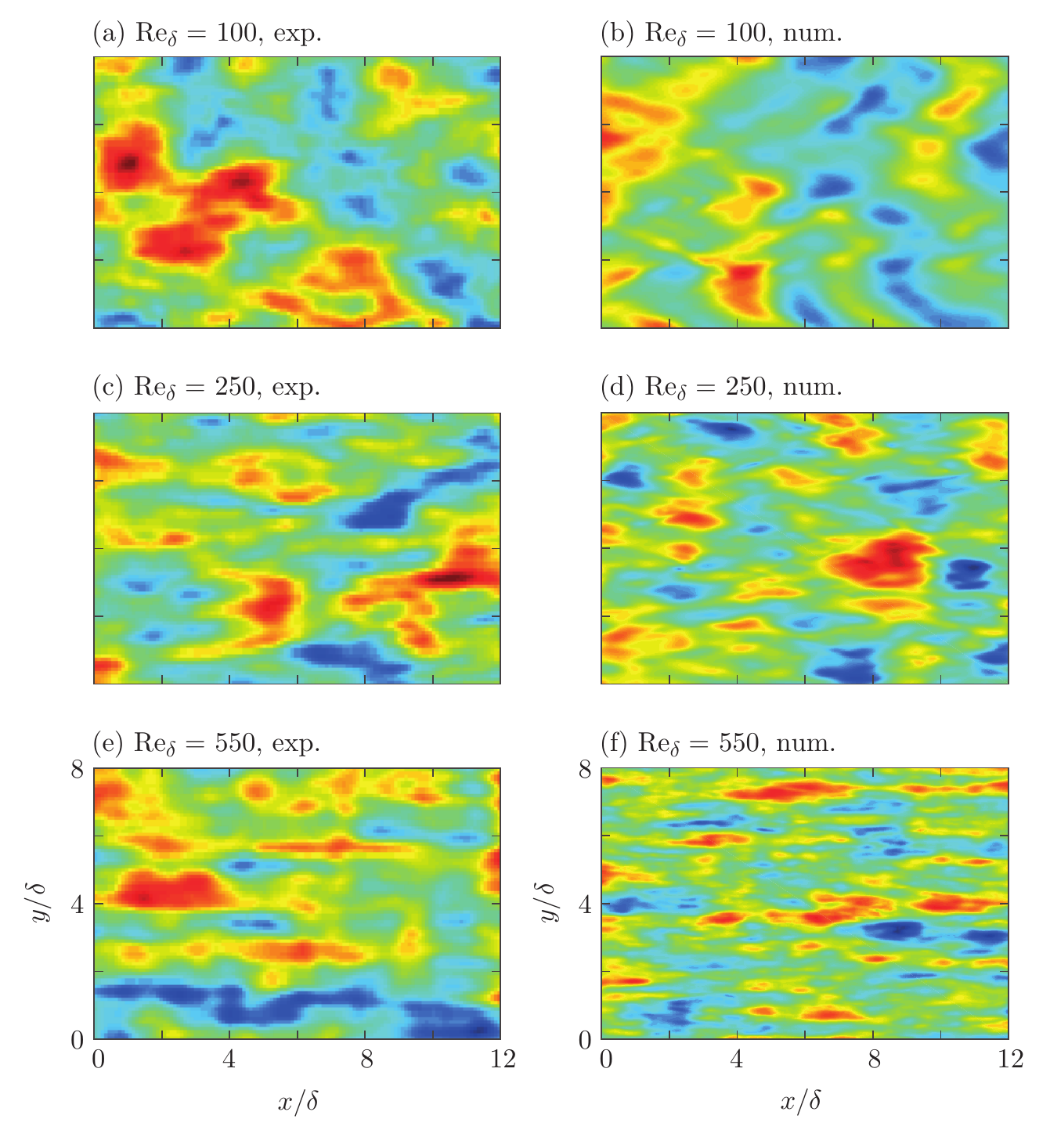}
\caption{Comparison between experimental surface displacement field $\zeta$ measured by~\citet{Paquier_2015} (left) and synthetic wrinkle fields computed from DNS data with Eq.~(\ref{eq_psimpreal}) (right). Only a subdomain of the synthetic wrinkle field is shown, to match the size of the experimental domain.  Experimental data (left) for $U_a = 1$\,m/s (a), 2.5 \,m/s (c), and 5.5 \,m/s (e), which corresponds to Re$_\delta$ = 100 (b),  Re$_\delta$ = 250 (d),  Re$_\delta$=550 (f).}
\label{fig4}
\end{figure}

Figure~\ref{fig4} shows snapshots of synthetic wrinkle fields computed using this direct method, for Re$_\delta = 100, 250$ and 550, compared to experimental measurements by \cite{Paquier_2015} for a liquid viscosity $\nu_{\ell}=30$~mm$^2$~s$^{-1}$ and equivalent Reynolds numbers (the corresponding wind speeds are $U_a$ = 1, 2.5 and 5.5\,m/s). These synthetic fields are obtained from the pressure and shear stress snapshots shown in figure~\ref{fig2}. The wrinkles appear as disordered fluctuations, nearly isotropic at Re$_\delta = 100$, that become elongated in the streamwise direction as Re$_\delta$ increases. We can note the good qualitative agreement between experimental and synthetic wrinkles for Re$_\delta=100$ and 250  (quantitative comparisons are provided in section 5.3). \textcolor{black}{By comparing figures~\ref{fig2} and \ref{fig4}, we note that the characteristic size of the wrinkles is always significantly larger than the size of the pressure and shear stress pattern from which they originate. This shift of the surface response towards larger scales is a key feature of the wrinkle regime. More precisely, the width of the wrinkles slightly decreases with Re$_\delta$, while their length remains nearly constant.} The match between experimental and synthetic wrinkles is not as good at larger Reynolds number, for Re$_\delta = 550$: as the Reynolds number is increased, the DNS resolves smaller scales, while the experimental measurements reach the limit of resolution at an intermediate spatial scale, smoothing off the small scales. The experimental wrinkles at large Reynolds number appear here as a coarse-grained version of the synthetic wrinkles. \textcolor{black}{Another possible source of visual discrepancy is the appearance of evanescent regular waves in the experiment, which were shown to coexist with wrinkles slightly below the wind threshold \citep{Paquier_2015}. These regular waves, which originate from an instability mechanism, cannot be captured by the present linear theory.}

\begin{figure}
\centering
\includegraphics[width=0.85\columnwidth]{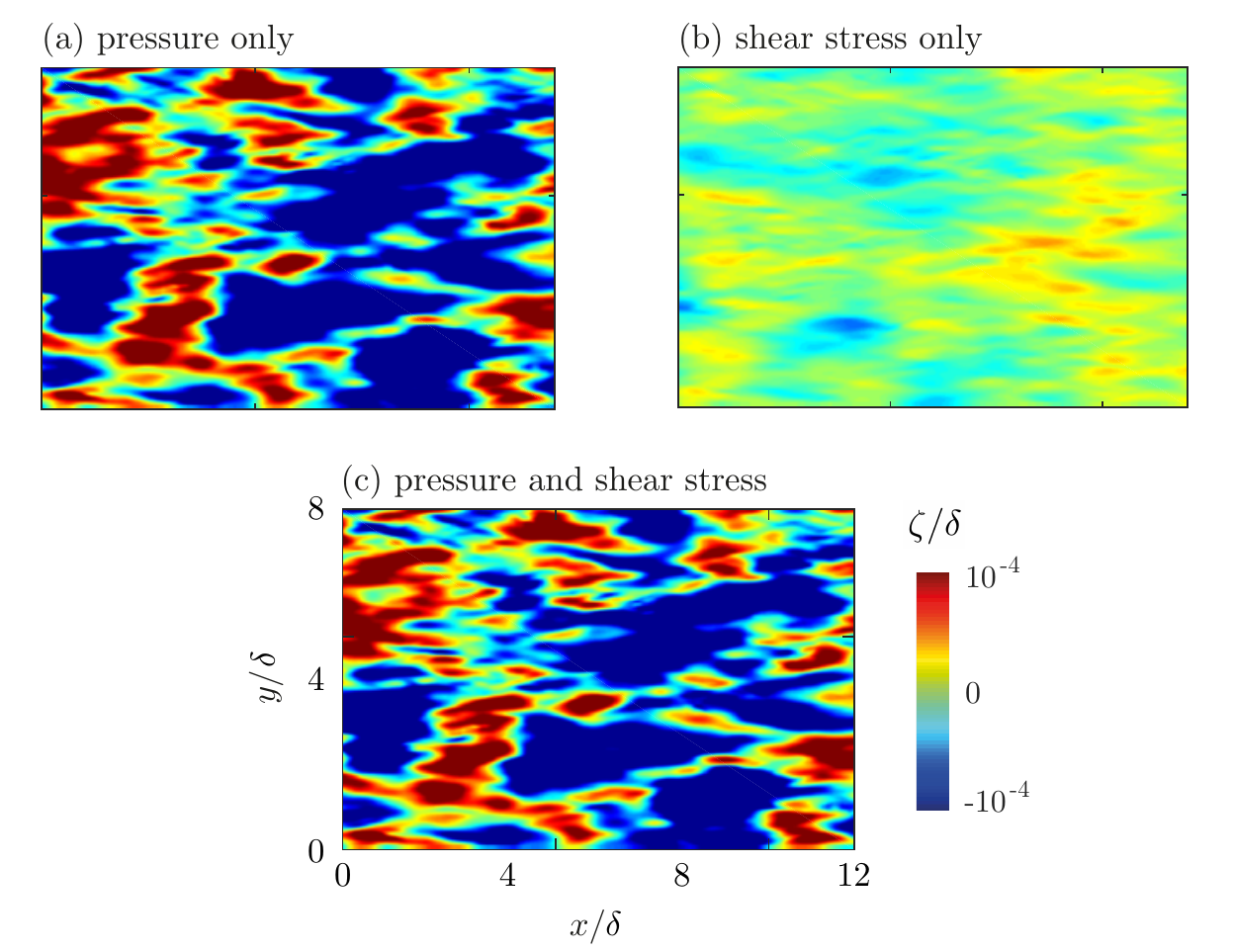}
\caption{Analysis of the respective role of pressure and shear stress fluctuations on the wrinkle generation.  Synthetic wrinkle field snapshot obtained for Re$_\delta = 250$ from (a) the pressure contribution $p$ only; (b) the shear stress contribution $\sigma$ only; (c) both pressure and shear stress. The full field is almost indistinguishable from the field (a), showing that the main contribution originates from pressure fluctuations.}
\label{fig4bis}
\end{figure}

An interesting question  is whether the wrinkles originate mostly from the pressure forcing or the shear stress forcing. The relative contribution of the two terms is illustrated in figure~\ref{fig4bis}, obtained at the intermediate Reynolds number Re$_\delta=250$. Figure~\ref{fig4bis}(a) and (b) show snapshots of surface deformation computed using the pressure contribution only ($\hat S^\dagger = \tilde k {\hat p}^+$) and shear stress contribution only ($\hat S^\dagger = i \tilde{\bs k} \cdot \hat{\bs \sigma}^+$), while figure~\ref{fig4bis}(c) combines the two contributions. Pressure clearly dominates the wrinkle generation. This important result can be primarily attributed to the larger rms amplitude of pressure (one has $p^+_\textrm{rms} \simeq 4 \sigma^+_\textrm{rms}$ here, see figure~\ref{fig3}a). However, this larger amplitude is not sufficient to explain the factor 10 in amplitude between figure~\ref{fig4bis}(a) and \ref{fig4bis}(b). The stronger influence of pressure also originates from the particular form of the transfer function $1/\hat D$ which tends to amplify structures of larger size. We can conclude that, although elongated in the streamwise direction, wrinkles are essentially disordered wakes generated by the nearly isotropic traveling pressure fluctuations.

\textcolor{black}{It is interesting to discuss the geometry of the wrinkles in the context of the Kelvin-Mach transition observed in ship wake patterns at large Froude number~\citep{Rabaud_2013,Darmon_2014}. We define the Froude number of a nearly isotropic pressure patch of characteristic dimension $\Lambda_x \simeq \Lambda_y \simeq 250 \, \delta \, \mathrm{Re}_\delta^{-1}$ traveling at velocity $U_c$ as $\mathrm{Fr} = U_c / \sqrt{g \Lambda}$. In figure~\ref{fig4}(a,b), one has $\mathrm{Fr} \simeq 0.7$, a value close to the transition $\mathrm{Fr}_c \simeq 0.5$ below which wakes are well described by the classical Kelvin wake pattern, of half-angle of $\alpha = \sin^{-1}(1/3) \simeq 19.5^\mathrm{o}$. Some oblique bands can indeed be distinguished in the snapshots, reminiscent of such Kelvin wakes. In figures~\ref{fig4}(b,c) and \ref{fig4}(d,e), one has $\mathrm{Fr} \simeq 2.8$ and 9 respectively, for which the wakes are in the Mach-like regime, characterised by a much smaller angle $\alpha \simeq 0.2~\mathrm{Fr}^{-1}$ ($\simeq 4.5^\mathrm{o}$ and $1.5^\mathrm{o}$, respectively). The thinning of the wrinkles at increasing $\mathrm{Re}_\delta$ is therefore a signature of the decreasing angle of the wakes generated by the traveling pressure patches. We finally note that the Bond number based on the size of the pressure patches, $\Lambda / \ell_c \simeq 250 \, \mathrm{Bo} \, \mathrm{Re}_\delta^{-1}$, decreases between 35 and 6 for the range of Reynolds numbers considered here. As observed in \cite{Moisy_PRE_2014}, the wakes are essentially in the gravity regime for these values, with weak capillary effects.}

\subsection{Analysis in Fourier space}

\begin{figure}
\centering
\includegraphics[width=0.9\columnwidth]{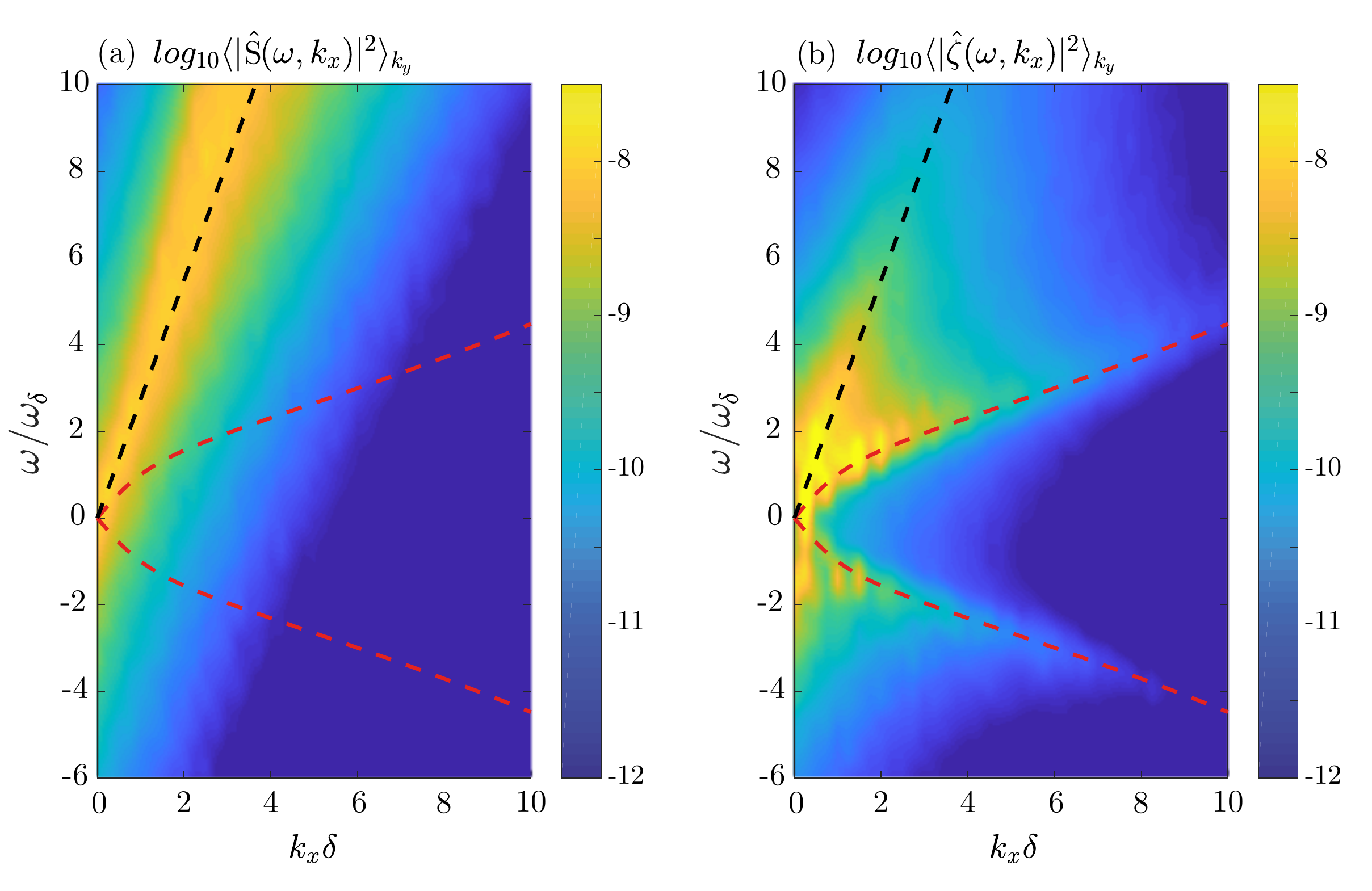}
\caption{(a) Space-time spectrum $\langle | \hat S|^2 \rangle_{k_y}$ of the source term for Re$_\delta = 250$. (b) Space-time spectrum of surface displacement $\langle | \hat \zeta(\omega,k_x) |^2 \rangle_{k_y}$, computed from Eq.~(\ref{eq_psimp}) for a liquid viscosity corresponding to $\nu_{\ell}=30$ mm$^2$/s.  Black dashed line: $\omega = k_x U_c$, where $U_c$ is the convection velocity. Red dashed line: dispersion relation.  \textcolor{black}{$k_x$ is normalised by the boundary layer thickness $\delta$, and $\omega$ is normalised by the frequency $\omega_\delta$ corresponding to a wave of wavenumber $k \delta = 1$.}}
\label{fig6}
\end{figure}

An important drawback of the direct reconstruction of synthetic wrinkles from the three-dimensional evaluation of Eq.~(\ref{eq_psimpreal}) is the strong discretisation effects for small $(\bs k, \omega)$. The most amplified Fourier modes in Eq.~(\ref{eq_pvol_dless}) occur in a narrow range of $(\bs k, \omega)$, where the transfer function $1/\hat D$ takes large values, that is difficult to resolve numerically on a discrete grid. Although the overall shape of the wrinkle fields are robust, their amplitude shows a significant dependence on the domain size and duration of the DNS, and hence on the discretisation in $(\bs k, \omega)$. To circumvent this discretisation issue, we perform a finer analysis of ${\hat S}^\dagger$ and $\hat \zeta$ in Fourier space, allowing for a refined and more robust evaluation of the wrinkle properties.

Figure~\ref{fig6} shows two-dimensional representations of the three-dimensional spectra $|\hat S(\bs k, \omega)|^2$ and $|\hat \zeta(\bs k, \omega)|^2$ computed from the DNS run at Re$_\delta = 250$, for a liquid viscosity corresponding to $\nu_\ell = 30$~mm$^2$/s in the experiment. The spectra are averaged along the spanwise direction, $k_y$, and plotted in the $(k_x, \omega)$ plane. By symmetry, only the two quadrants corresponding to $k_x>0$ are shown. \textcolor{black}{The wavenumber $k_x$ is normalised by the boundary layer thickness $\delta$, and the angular frequency is normalised by $\omega_\delta$, the angular frequency of a wave of wavenumber $k \delta = 1$. In Figure~\ref{fig6}(a), the energy of the source $\hat S$ is spread over a broad band centred around the line $\omega = k_x U_\mathrm{c}$ (black dashed line), where the convection velocity $U_\mathrm{c}$ corresponds to that measured in figure~\ref{fig3}(c). The width of the band is related to the correlation time of the turbulent fluctuations. A rigid pattern traveling at constant speed would correspond to a perfect accumulation of energy along the  line $\omega = k_x U_\mathrm{c}$.  The dispersion relation $\omega(k_x)$ is also plotted (red dashed line), showing that the forcing energy is mostly supplied to waves in the gravity regime (the capillary regime starts at $k_x \delta = Bo =14$, which is outside the axis of the figure).}

\begin{figure}
\centering
\includegraphics[width=1\columnwidth]{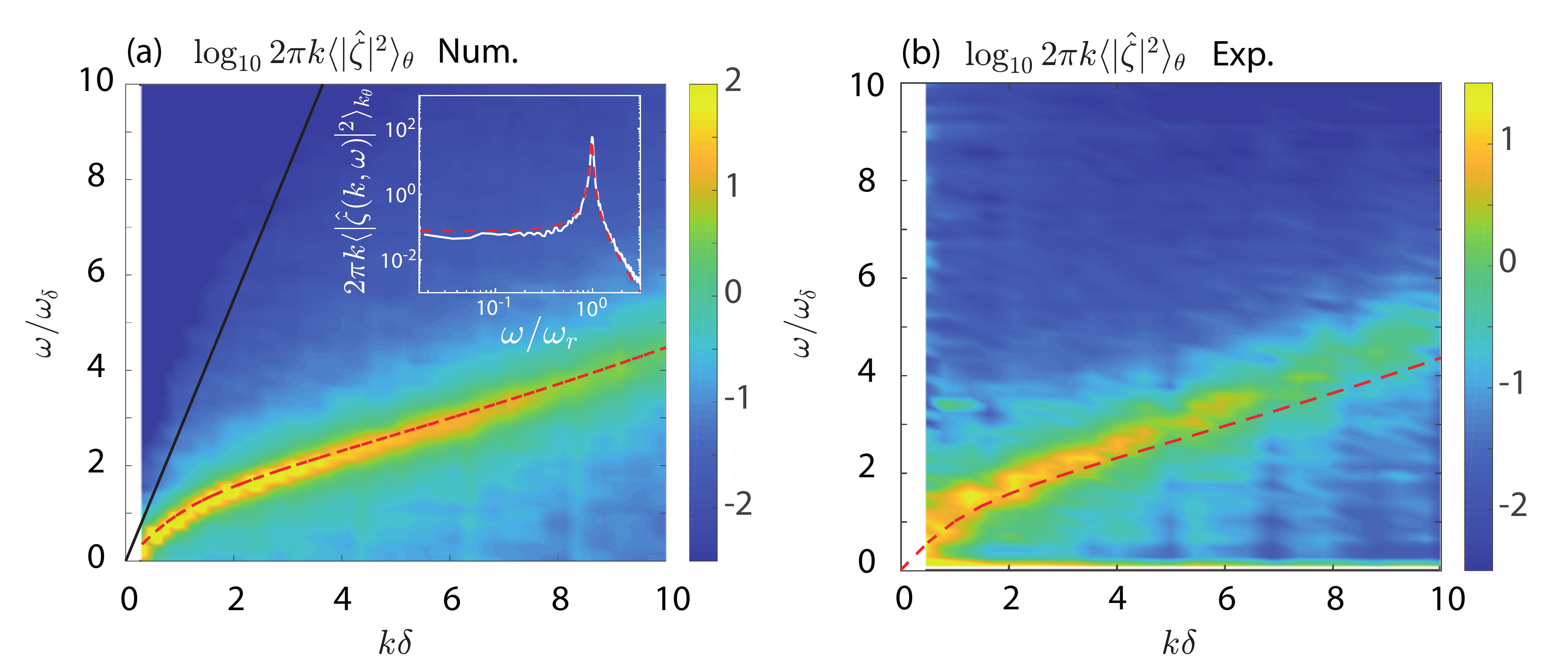}
\caption{(a) Space-time spectrum of surface displacement $\langle | \hat \zeta |^2 \rangle_{\theta}$ averaged over the azimuthal direction for $\nu_{\ell}=30$ mm$^2$/s, computed from Eq.~(\ref{eq_psimp}) with the DNS data at Re$_\delta=250$. The three dimensional spectrum is averaged along the direction $\theta$ to respect the symmetry of the dispersion relation.  \textcolor{black}{The black line shows the forcing $\omega = U_c k$, and the red dashed line is the dispersion relation. Inset: Spectrum for a given wavenumber $k \delta = 2.5$ computed from DNS (solid white line), and its Taylor expansion (\ref{eq_tayl}) around the corresponding resonant frequency $\omega_r$ (red dashed line). The peak half-width is given by the viscous frequency, $\omega_\nu = 4 \nu_\ell k^2 \simeq 0.03 \omega_\delta$.} (b) Space-time spectrum $\langle | \hat \zeta|^2 \rangle_{\theta}$ computed from the experimental data from~\citet{Paquier_2015}, for $U_a$ = 2.5\,m/s and $\nu_{\ell}=30$\,mm$^2$/s. }
\label{fig6bis}
\end{figure}

Representing the three-dimensional spectrum $\hat \zeta (\bs k, \omega)$ in a two-dimensional form is delicate, because of the lack of symmetry of  Eq.~(\ref{eq_psimp}) in the plane $(k_x,k_y)$. We provide in figure~\ref{fig6}(b) a two-dimensionnal representation of the spectrum, $\langle |\hat \zeta|^2 \rangle_{k_y}$, using an averaging along $k_y$ as for the source $\hat S$. We see that the energy of the surface response is located at smaller wavenumbers than the forcing, confirming that the wrinkles are of larger size than the pressure and shear stress structures.  The resonant response is mostly contained in the $\omega>0$ quadrant, but it also has a significant amount of energy in the $\omega<0$ quadrant, indicating a small counter-propagating component. In that representation, energy accumulates around two regions: a first region surrounding the dispersion relation (red dashed line), and a second region surrounding the forcing $\omega = U_{\mathrm{c}} k_x$ (black dashed line).  This second region would suggest that a significant amount of energy is away from the resonance. However, this apparent non-resonant response is an artifact of the averaging over $k_y$ which respects the symmetry of the source but not that of the dispersion relation (see sketch of figure~\ref{fig1}). This bias is removed in figure~\ref{fig6bis}(a), showing the same spectrum now averaged in the azimuthal direction,  $\langle |\hat \zeta^2| \rangle_{\theta}$, as a function of $k = (k_x^2+k_y^2)^{1/2}$. In that representation, all the energy is now located near the dispersion relation (red dashed line). This clearly indicates that the accumulation of energy along the forcing $\omega = U_{\mathrm{c}}k_x$ in figure~\ref{fig6}(b) was a contribution of the Fourier components satisfying the dispersion relation with $k_y \neq 0$.

The strong accumulation of energy along the dispersion relation is also present in the experiment:  Figure~\ref{fig6bis}(b) shows the azimutally averaged spectrum $\langle |\hat \zeta^2| \rangle_{\theta}$ computed from the experimental surface deformation fields for the same Reynolds number Re$_\delta = 250$. In spite of the lower spatial resolution of the experimental data, a clear accumulation of energy appears in the vicinity of the dispersion relation. We can note a slight shift of energy at frequency larger than the dispersion relation. The shift may be attributed to the surface drift current $U_\mathrm{s}$, which yields a Doppler-shifted dispersion relation $\omega \simeq g' k \tanh(k h) + U_\mathrm{s} k_x$.

\textcolor{black}{The dominant response along the dispersion relation, observed both experimentally and numerically, confirms that the wrinkles are a superposition of a broad range of propagating waves. Their main specificity is their non trivial transverse structure: While regular waves correspond to $k_y=0$, wrinkles are characterised by wave vectors ${\bs k}$ tilted with respect to the wind direction ($k_y \neq 0$), leading to elongated patterns, statistically stationary along $y$ (because of the symmetry $k_y \rightarrow -k_y$) and propagating along $x$.}

\subsection{Integration of the resonant response}

Since the main surface response occurs along the dispersion relation, we may simplify further the three-dimensional integral (\ref{eq_psimpreal}) by considering only the resonant response. In the limit of narrow resonance ($\omega_\nu \ll \omega_r$, i.e. $\tilde \omega_r \gg 1$) and of slow varying source amplitude over the width of the resonance ($\omega_\nu \partial_\omega |\hat S|^2 \ll |\hat S(\bs k,\omega_r)|^2$), the integrand of Eq.~(\ref{eq_pvol_dless}) may be substituted by its second-order Taylor expansion,
\bb
\frac{|{\hat S^\dagger(\tilde{\bs k},\tilde \omega)}|^2}{(\tilde \omega-\tilde \omega_r)^2 (\tilde \omega / \tilde \omega_r +1)^2 + (\tilde \omega / \tilde \omega_r)^2} &\simeq& \frac{|{\hat S^\dagger(\tilde{\bs k},\tilde \omega_r)}|^2}{1 + 2 (\tilde \omega - \tilde \omega_r) + (4+\tilde \omega_r^{-2}) (\tilde \omega-\tilde \omega_r)^2 }\\ \nonumber
 &+& O((\tilde \omega- \tilde \omega_r)^{-2}) .
 \label{eq_tayl}
\ee
Note that the odd term $(\tilde \omega - \tilde \omega_r)$ of the expansion does not contribute to the wrinkle amplitude as it cancels out upon integration along $\omega$.

The validity of this approximation is illustrated in the inset of figure~\ref{fig6bis}(a): the compensated spectrum $2 \pi k \langle |\hat \zeta(k,\omega)|^2 \rangle_{\theta}$, plotted as a function of the angular frequency $\omega$ for a fixed wavenumber, $k \delta = 2.5$ (white curve), is indeed very close to its second order Taylor expansion (red dashed line).   The limit of narrow resonance breaks down for $\omega_\nu \approx \omega_r$, which would occur around $k \delta = 10$ for viscosity $\nu_\ell = 30 \nu_{\textrm{water}}$. In practice, most of the energy is contained in the region $k \delta < 10$, so  that the full integral of Eq.~(\ref{eq_pvol_dless}) can be safely approximated by the limit of narrow resonance.

\begin{figure}
\centering
\includegraphics[width=0.95 \columnwidth]{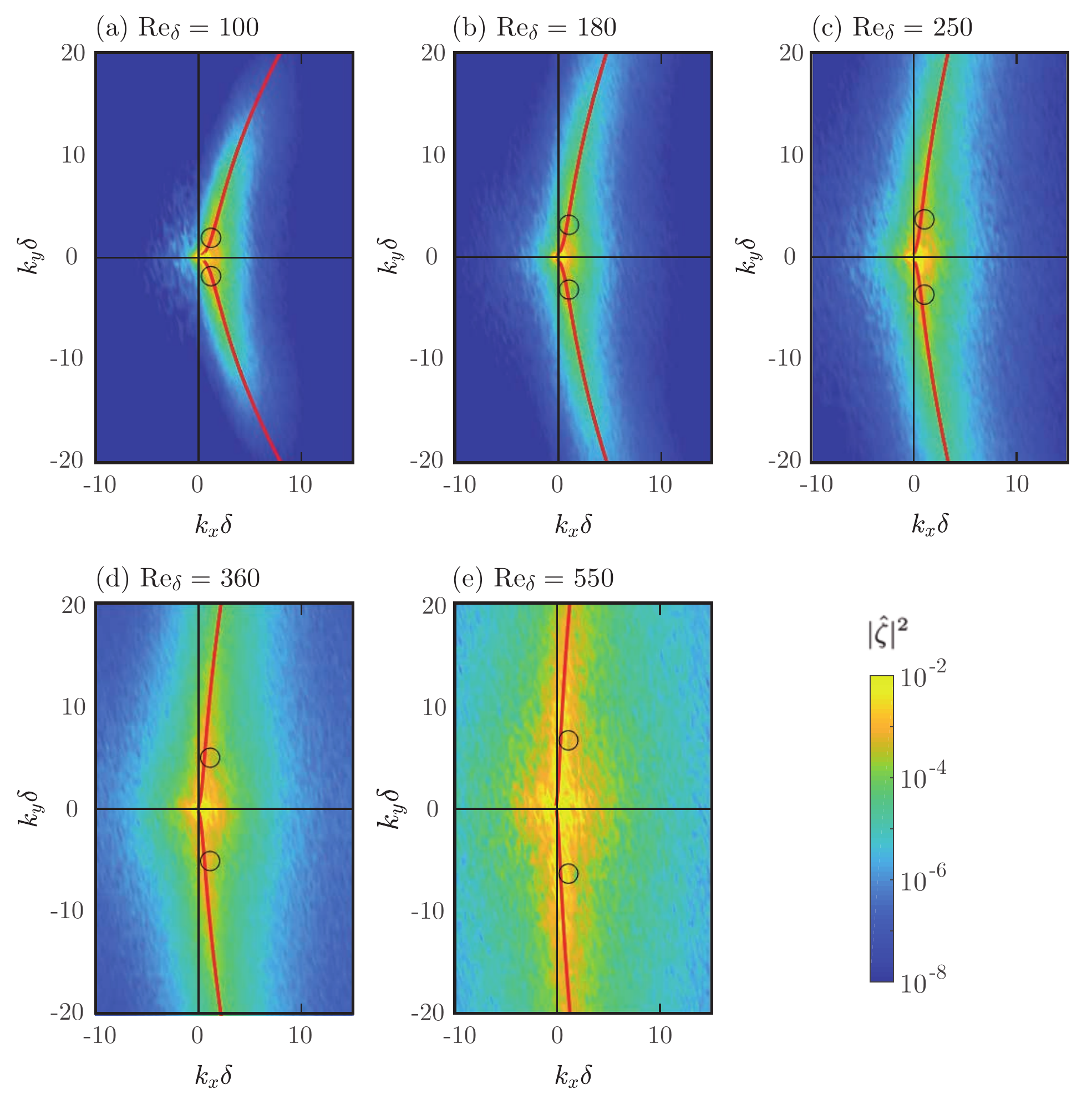}
\caption{Spectrum  of the surface displacement $|\hat  \zeta|^2$ integrated along the dispersion relation for Re$_\delta$ = 100, 180, 250, 360 and 550. The resonant curve, defined as the intersection between the forcing plane $\omega = k_x U_\mathrm{c}$ and the surface $\omega_r = g' k \tanh( kh)$, is shown in red line. The spectral barycenter ($K_x,K_y$) is shown ($\circ$) for each quadrant $k_x>0$.}
\label{fig10}
\end{figure}

In that limit, the integral along $\omega$ in Eq.~(\ref{eq_pvol_dless}) can be performed analytically, and we obtain
\bb
\langle | \hat \zeta^2 | \rangle_{\omega} ({\bs k}) =  \int {\rm d}\omega |\hat \zeta({\bs k},\omega)|^2 =   \left (\frac{\rho_\mathrm{a}}{\rho_\ell} \right )^2 \frac{{u^*}^3}{16 g \nu_\ell} \frac{1}{(2\pi)^2}  W (\tilde{\bs k}) \, |\hat S^\dagger(\tilde{\bs k},\tilde \omega_r)|^2,
\label{zetaint}
\ee
where we introduce the weighting factor
\bb
W(\tilde{\bs k}) = \frac{1}{\tilde k^3 (1+ \mathrm{Bo}^{-2} \tilde k^2 ) \tanh (\tilde k h/\delta )}.
\label{eq_wf}
\ee
This weighting factor originates from the expression of the dispersion relation, and is responsible for the shift towards larger scales of the surface response. This wavelength shift depends on the values of the Bond number Bo and the dimensionless depth $h/\delta$. For $k \ll 1/h$, the waves are in a shallow water regime, and the weighting factor is $W(\tilde k) \simeq \delta/(h \tilde k^4)$. For $1/h \ll k \ll 1/\ell_c$, the waves are in the gravity regime and  $W(\tilde k) \simeq 1/\tilde k^3$. Finally, for $k \gg 1/\ell_c$, the waves are in the capillary regime and $W(\tilde k) \simeq \mathrm{Bo}^{2} / \tilde k^5$. In all cases, this weighting factor tends to enhance the low$-k$ content of the forcing. In practice, most of the energy is contained in the deep-water gravity regime, so that $W(\tilde k) \simeq 1/\tilde k^3$ is the most relevant weighting factor in the wrinkle problem.

From Eq.~(\ref{zetaint}), the surface response can be computed using a linear interpolation of $\hat S^\dagger({\bs k},\omega)$ on the dispersion relation manifold $\hat D({\bs k},\omega)=0$. The interpolation is performed along the direction $\omega$ using 20 mesh points in the range $[0.8 \omega_r,1.2\omega_r ]$ and the same mesh size in $k$ than the original mesh size, and the value at the resonance is taken as the average over the 20 computed values of $\omega$ for each $\bs k$.   This procedure reduces the computational cost by a factor up to 100 for the highest Reynolds number. Its main advantage is to remove the evaluation of the full three-dimensional integral which presents a highly peaked resonance near $\omega = \omega_r$. As a consequence, the properties of the wrinkles are found to be less sensitive to the spatial and temporal discretisation of the DNS data.

\begin{figure}
\centering
\includegraphics[width=1. \columnwidth]{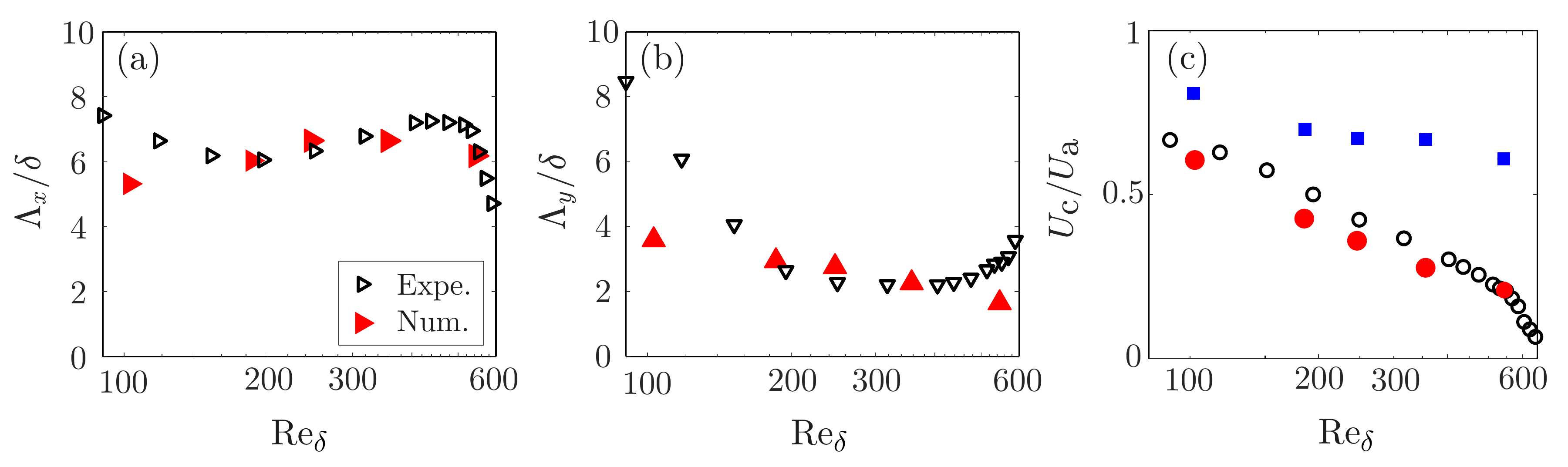}
\caption{Sizes and convection velocity of the wrinkles as a function of Re$_\delta$. Comparison between the synthetic wrinkles (red filled symbols) and the experimental wrinkles of~\citet{Paquier_2015} (black open symbols).  (a) Streamwise size $\Lambda_x/\delta$, (b) spanwise size $\Lambda_y/\delta$, and (c) convection velocity $U_\mathrm{c}/U_{\mathrm{a}}$ computed from Eqs.~(\ref{defK}), (\ref{defO}) and (\ref{defUc}). The blue squares show the convection velocity of the pressure forcing (see figure~\ref{fig3}(c)).}
\label{fig11}
\end{figure}

Figure~\ref{fig10} shows the Fourier spectrum of the surface deformation (\ref{zetaint}) for increasing values of the Reynolds number Re$_\delta$. The energy is essentially contained in a  zone located at relatively small $k$, with a symmetric distribution along $k_y=0$.  It essentially follows the intersection of the dispersion relation manifold and the forcing plane $\omega = k_x U_{\mathrm{c}}$, given by
\bb
g k \tanh(k h) (1+(k \ell_c)^2) = (k_x U_{\mathrm{c}})^2,
\ee
plotted in red line using for each $Re_\delta$ the convection velocity measured in figure \ref{fig3}(c). {\textcolor{black}While the energy remains mostly located in the range $k_x \delta \in [0,2]$, it significantly broadens in the $k_y$ direction as $Re_\delta$ increases, i.e. the angle of the phase velocity gradually departs from the direction of the wind. This increase of $k_y / k_x$ is the spectral signature of the thinning  of the wrinkles in the spanwise direction.}  More specifically, we can compute the spectral barycenter $\bs K$ of 
$|\hat \zeta|^2$ from Eq.~(\ref{defK}), shown as circles in figure~\ref{fig10}.  The corresponding streamwise and spanwise sizes, defined as $\Lambda_x = 2 \pi /K_x$ and $\Lambda_y = 2 \pi / K_y$, are plotted in figure~\ref{fig11}(a,b), and compared to the experimental data of \cite{Paquier_2015}. The almost constant streamwise size $\Lambda_x/\delta = 6.7 \pm 0.7$ and the decreasing spanwise size $\Lambda_y/\delta$ are qualitatively recovered. {\textcolor{black}It is worth noting that, in spite of their large {\it streamwise} extent compared to the liquid depth, wrinkles are essentially deep-water waves: their wavenumber $k$ is dominated by the {\it spanwise} component $k_y$, for which we have $\tanh(kh) > 0.97$ in this range of Re$_\delta$. Last but not least, the computed convection velocity, shown in figure~\ref{fig11}(c), closely follows the experimental data: at small $Re_\delta$, the normalized convection velocity $U_\textrm{c} / U_\textrm{a}$ of the wrinkles is close to that of the pressure forcing ($U_\textrm{c} / U_\textrm{a}\simeq 0.7$, blue squares), but it decreases significantly as Re$_\delta$ increases.  This decreasing convection velocity for the wrinkle is a consequence of the propagation angle of the Fourier modes composing the wrinkles, which gradually departs from the direction of the wind as Re$_\delta$ increases.}

The correct agreement between synthetic and experimental wrinkles can be extended to the range of viscosity $\nu_{\textrm{water}}< \nu_\ell < 10^3 \nu_{\textrm{water}}$, since both the experimental measurement and the theory do not exhibit variation in viscosity on sizes and convection speed. At large Reynolds number, Re$_\delta>400$, the sharp decrease in $\Lambda_x$ and increase in $\Lambda_y$ found experimentally is not reproduced by the computation. This sharp evolution is associated to the instability that gives rise to the regular waves, which cannot be captured by the present linear model. 

\begin{figure}
\centering
\includegraphics[width=0.8 \columnwidth]{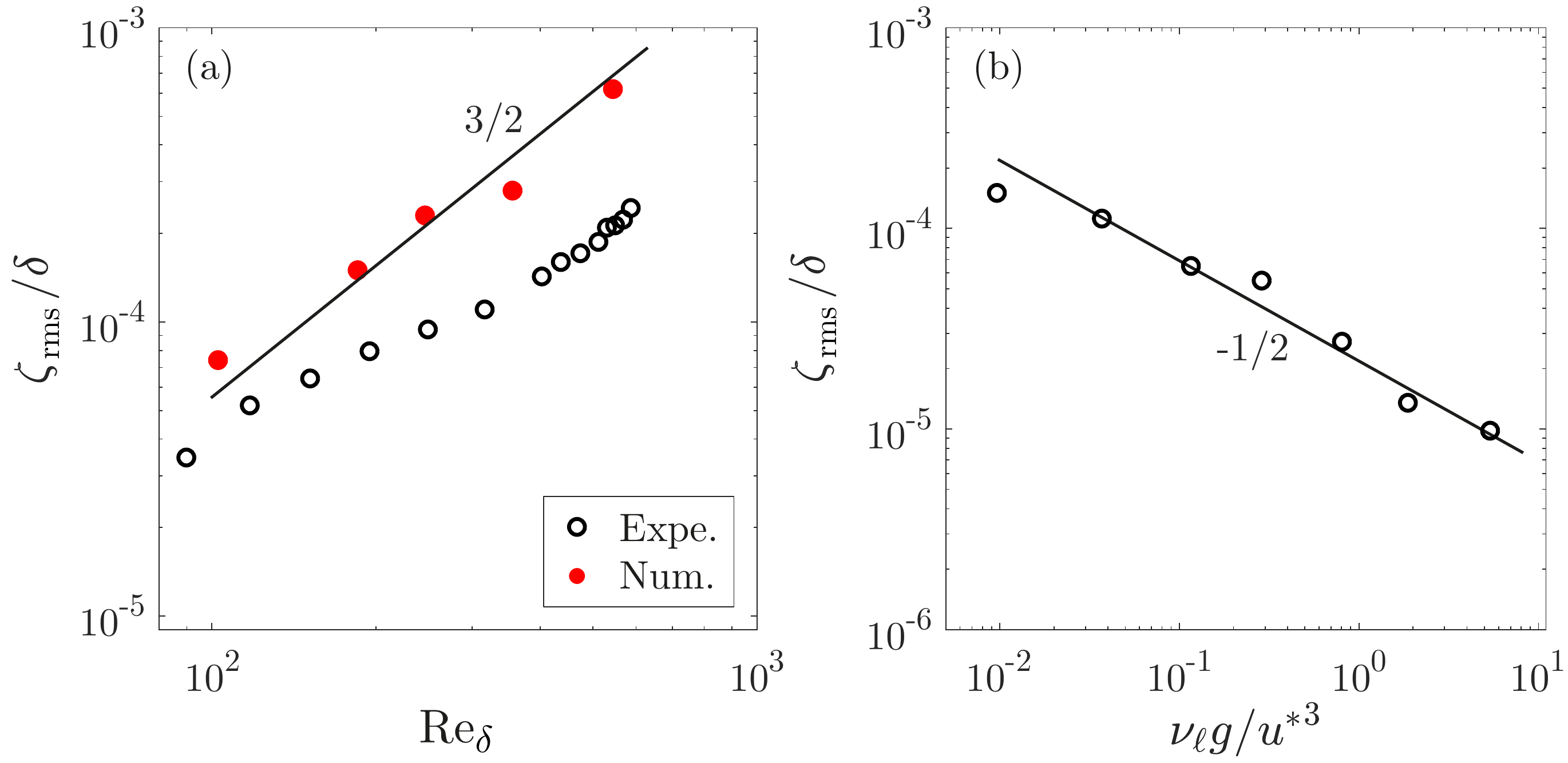}
\caption{Scaling of the wrinkle amplitude $\zeta_{\textrm{rms}}/\delta$ with respect the Reynolds number Re$_\delta$ and the liquid viscosity $\nu_\ell$.  ($\circ$) experimental data; ({\red $\bullet$}) synthetic wrinkles. (a) $\zeta_{\textrm{rms}}/\delta$ as a function of Re$_\delta$ for $\nu_\ell = 30~$mm$^2$/s. (b) $\zeta_{\textrm{rms}}/\delta$ as a function of the dimensionless number $\nu_\ell g/ {u^*}^3 $ for $Re_\delta=180$ and liquid viscosity in the range $1 - 600$~mm$^2$~s$^{-1}$.  A best fit with the analytical prediction (\ref{eq_zfinalC}) is shown in black solid line, yielding $C \simeq 0.022$.}
\label{fig12}
\end{figure}

We finally turn to the scaling of the wrinkle amplitude as a function of the friction velocity $u^*$ and liquid viscosity $\nu_\ell$. By considering only the resonant contribution, the wrinkle amplitude $\zeta_{rms}$ is obtained from Eq.~(\ref{zetaint}) by summing $\langle | \hat \zeta^2 | \rangle_{\omega}$ over ${\bs k}$,
\textcolor{black}{
\bb
\zeta_{rms}^2 =   \left (\frac{\rho_\mathrm{a}}{\rho_\ell} \right )^2 \frac{{u^*}^3}{16 g \nu_\ell} \frac{1}{(2 \pi)^2} \int \mathrm {\rm d}^2 \tilde{\bs k} \ W(\tilde{\bs k}) \ |\hat S^\dagger(\tilde{\bs k},\tilde \omega_r)|^2 .
\label{zetar}
\ee
}
Figure~\ref{fig12}(a) shows the wrinkle amplitude $\zeta_{\textrm{rms}}/\delta$ as a function of Re$_\delta$. The experimental scaling in $\zeta_{\rm rms} \propto {u^*}^{3/2}$ is well reproduced by the the synthetic wrinkles, but with an amplitude twice larger. This discrepancy probably originates from the high sensitivity to the low-wavenumber content of the forcing. The contribution of the largest scales to the wrinkle amplitude $\zeta_{\rm rms}$ can be different experimentally and numerically for two reasons. First, the measurements were carried out on a window smaller than the channel width, so that the smallest wavenumbers may be poorly estimated. Second, the numerical simulations are performed in a box with periodic boundary conditions, so the largest scales could be different from that of a true developing turbulent boundary layer.

Finally, we show in figure~\ref{fig12}(b) the experimental wrinkle amplitude $\zeta_{rms}$ as a function of the dimensionless liquid viscosity $\nu_\ell g/ {u^*}^3$ for a fixed value of $u^*$ corresponding to Re$_\delta = 180$. The data are in good agreement with the analytical prediction $\zeta_{\rm rms}/\delta \propto (\nu_\ell g/{u^*}^3)^{-1/2}$.  We can conclude that the dimensionless function $f_4$ introduced in Sec.~2.1 is essentially independent of the Reynolds number in the range Re$_\delta \in [100,550]$. The additional dependencies in Bo and $h/\delta$, explicitly considered in the derivation, cannot be tested against experiments, which were performed for fixed surface tension and liquid depth (Bo~$ \simeq 14$ and $\delta/h \simeq 1.2$, see section 2.2). Ignoring these dependencies in Bo and $h/\delta$, the function $f_4$ reduces to a constant, and Eq.~(\ref{eq:f4}) simply writes:
\bb
\frac{\zeta_{\mathrm{rms}}} {\delta} \simeq C \frac{\rho_a}{\rho_\ell} \frac{{u^*}^{3/2}}{(g \nu_\ell)^{1/2}}.
\label{eq_zfinalC}
\ee 
The numerical factor, fitted in figure~\ref{fig12}(b), is $C \simeq 0.022 \pm 0.005$.

\section{Connection with the inviscid resonant theory of \citet{Phillips_1957}}

The present theory focuses on the statistically steady wrinkle regime, of amplitude governed by the liquid viscosity. This regime is the asymptotic state of the surface deformation, reached when the energy input by the turbulence forcing is balanced by the viscous dissipation. Before this energy balance is reached, a transient growth regime must take place, where viscous dissipation can be neglected. We show here how this inviscid growth regime, previously investigated by \citet{Phillips_1957}, naturally asymptotes towards the viscous saturated wrinkle regime described in the present theory, provided that the wrinkles remain of small amplitude.

\subsection{Temporal growth of wrinkles}
	
\begin{figure}
\centering
\includegraphics[width=0.75 \columnwidth]{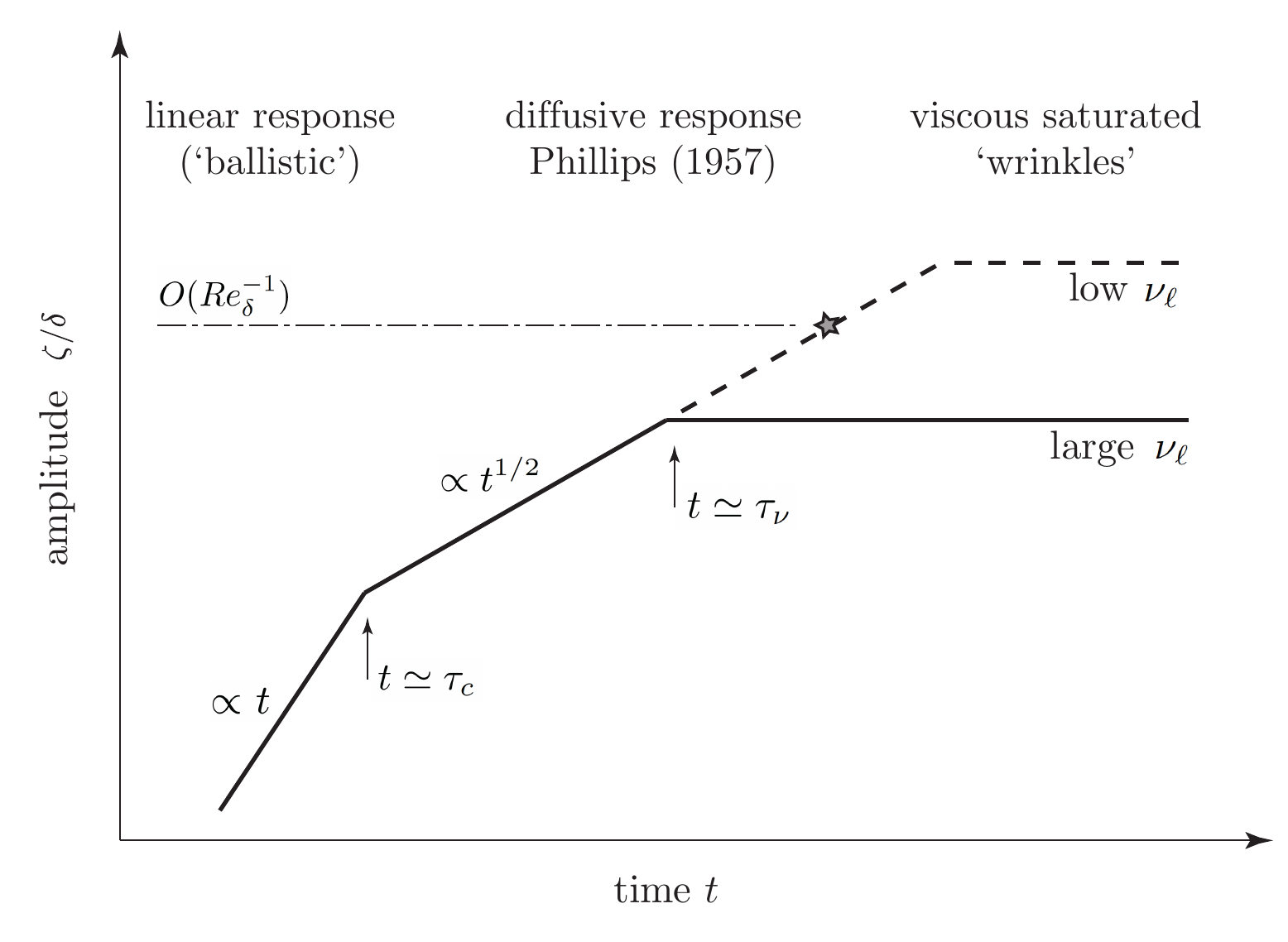}
\caption{Sketch of the surface deformation amplitude $\zeta/\delta$ as a function of time for a given $Re_\delta$ and two liquid viscosities $\nu_\ell$. Starting from a purely flat interface, three regimes follow one another: linear response at short time, for $t \ll \tau_c$, where $\tau_c \simeq \delta_v / u^*$ is the correlation time of the turbulent structures; Phillips (1957) regime of quasi-inviscid resonant growth at intermediate time, for $\tau_c \ll t \ll \tau_\nu$, where $\tau_\nu \simeq \nu_\ell / \delta^2$ is the viscous timescale; asymptotic viscous-saturated wrinkle regime at long time for $t \gg \tau_\nu$. If the wrinkle amplitude reaches a given fraction of the viscous sublayer thickness (star symbol, for $\zeta/\delta \simeq Re_\delta^{-1}$) before the viscous saturation, the linear assumption of the present theory breaks down, resulting in the possible triggering of a wave instability.}
\label{fig_sketch_growth}
\end{figure}

In the previous sections, all fields were assumed statistically homogeneous and stationary, allowing for a space-time Fourier description. To describe the transient growth that precedes this steady regime, we still assume here homogeneity but we relax the stationary assumption. Only a spatial Fourier transform of the dynamical equation is then performed:
\begin{subeqnarray}
\breve \zeta({\boldsymbol k},t) &=& \int{\rm d}^2 \boldsymbol r \zeta({\boldsymbol r},t) e^{-i\bs k \cdot \bs r} \\
\zeta({\boldsymbol r},t) &=& (2\pi)^{-2} \int{\rm d}^2 \bs k  \breve \zeta({\boldsymbol k},t) e^{i\bs k \cdot \bs r} \ ,
\label{zetafouriert}
\end{subeqnarray}
noting $\breve \zeta$ the spatial Fourier transform for $\zeta$ and similarly for the pressure and stress fields. The dynamics is now governed by a Langevin equation~\textcolor{black}{\citep{Langevin_1908,Pottier_2014}} for the stochastic wave amplitude $\breve \zeta$, which can be derived following an approach similar to the derivation of section 3,
\bb
\partial_{tt} \breve \zeta(\bs k,t) + 4 \nu_\ell k^2 \partial_t \breve \zeta(\bs k,t) + g' k \breve \zeta(\bs k,t) = -k \breve p_0(\bs k,t) - i \bs k \cdot \breve{\bf \sigma}_0(\bs k,t).
\label{eq_ptime}
\ee
Each Fourier component $\bs k$ describes a linear damped oscillator forced by a stochastic noise given by the corresponding Fourier component of the applied pressure and shear stress fields. Such Langevin equation with short-time temporal correlations in the noise term exhibits three regimes, sketched in figure~\ref{fig_sketch_growth}: ballistic motion at short time ($\breve \zeta \propto t$), diffusive process at intermediate time ($\breve \zeta \propto t^{1/2}$), and asymptotic regime governed by cumulative effect of viscosity at large time ($\breve \zeta \propto t^0$). The intermediate-time regime, defined only for liquids of small viscosity, corresponds to the inviscid resonant theory of \citet{Phillips_1957}, whereas the large-time saturated regime corresponds to the wrinkles.

More specifically, we can introduce for each Fourier component $\bs k$ a fast correlation time $\tau_c(k) \sim (\bs k \cdot {\bf U}_{\bs k})^{-1}$, characterising the temporal correlation of the turbulent structures, and a slow dissipation time $\tau_\nu(k) \sim (\nu_\ell k^2)^{-1}$, associated to viscous dissipation in the liquid. We assume here for simplicity that the convection velocity ${\bf U}_{\bs k}$ is the same for all $\bs k$, and given by the global convection velocity $U_c \, {\bf e}_x$. In the intermediate-time regime $\tau_c \ll t \ll \tau_\nu$, each mode ${\bs k}$ corresponds to an essentially undamped oscillator forced by an uncorrelated noise, resulting in a linear growth of $|\breve \zeta (\bs k,t)|^2$, with a $\bs k$-dependent growth rate governed by the corresponding Fourier component of the pressure forcing (the shear stress forcing may be ignored during this quasi-inviscid growth).  The mean square wave amplitude, integrated over all modes, similarly grows linearly in time, resulting in the classical result (\ref{eq_phillips}) of \citet{Phillips_1957}. For time larger than the slowest (largest-scale) growing mode $\tau_\nu \simeq \delta^2 / \nu_\ell$, all Fourier components are saturated, and the asymptotic mean square amplitude can be simply estimated by setting $t \simeq \tau_\nu$ in Eq.(\ref{eq_phillips}): with $\overline{p^2} \propto \rho_a^2 u^{*4}$ and $U_c \propto u^*$, we recover the result of Eq.~(\ref{eq_zfinalC}). The wrinkle regime described in this paper therefore naturally arises as the viscous-saturated asymptotics of the inviscid growth theory of \citet{Phillips_1957}.

\subsection{Onset of regular waves}
\label{sec:orw}

A key assumption in our theory, as well as in the inviscid resonant theory of \citet{Phillips_1957}, is the absence of feedback of the wrinkles deformations on the turbulence in the air. In other words, the growth and saturated regimes sketched in figure~\ref{fig_sketch_growth} hold only provided that the wrinkle amplitude remains small compared to the thickness of the viscous sublayer $\delta_\nu$ in the turbulent air flow: this is assumption (g), used to derive the linearised surface response. The questions that naturally arise now are what is the maximum wrinkle amplitude before the breakdown of this assumption, and whether this breakdown could be related to the onset of regular waves. 

The breakdown of the decoupled dynamics hypothesis (assumption (g)) can be expected when the amplitude of the wrinkles reaches a fraction of the viscous sublayer thickness, $\zeta \simeq A \delta_\nu$, with $\delta_\nu = \nu_a / u^*$ and $A$ a numerical factor. Using Eq.~(\ref{eq_zfinalC}), this criterion is satisfied for a friction velocity $u^*$ beyond a critical value,
\bb
u^*_c = \left( \frac{A}{C} \right)^{2/5} \left( \frac{\rho_\ell}{\rho_a} \right)^{2/5} \left( \frac{g \nu_\ell \nu_a^2}{\delta^2} \right)^{1/5}.
\label{eq_ucstar}
\ee
The dependence of $u_c^*$ with liquid viscosity turns out to be remarkably close to the empirical law for the onset of regular (quasi-monochromatic) waves found in \citet{Paquier_2016},
\bb
u_c^* \simeq (2.3 \pm 0.2) \nu_\ell^{0.20}
\label{eq_ucstaremp}
\ee
($u_c^*$ in m/s,  $\nu_\ell$ in m$^2$/s). Identifying the numerical factor in (\ref{eq_ucstar}) from the empirical law (\ref{eq_ucstaremp}) yields $A \simeq 0.11 \pm 0.02$. This good match with the scaling $\nu_\ell^{1/5}$ suggests that regular waves could be triggered by an instability originating from the feedback of the wrinkles on the air turbulence: once the wrinkle amplitude reaches $A \delta_\nu$, the pressure and shear stress fluctuations in the boundary layer are no longer that of a no-slip flat surface, but acquire a spatio-temporal structure reflecting the shape of the surface. In turn, this spatio-temporal phase coherence between the wave field and the forcing could enhance the energy transfer, leading to the exponentially growing waves found in experiments. In this scenario, wrinkles appear as the natural base state from which regular waves grow. This scenario cannot be tested by the present theory, which ignores such coupling between the liquid and the air phases.

It may be noted that this tentative criterion for wave onset suggests that the turbulent boundary layer becomes sensitive to the surface roughness for rms amplitude of order of $0.1 \delta_\nu$. Such roughness is surprisingly small: the peak of turbulent kinetic energy in a boundary layer is at $15 \delta_\nu$, and the boundary layer is essentially a laminar shear flow up to $4\delta_\nu$. Boundary layer turbulence over a wavy no-slip wall is indeed essentially unaffected by rigid wall roughness up to $\simeq 4 \delta_\nu$~\citep{Schlichting_2000,Jimenez_2004}. The relatively small wrinkle amplitude found here for the growth of regular waves probably originates from the specific phase coherence of the surface waves and the pressure perturbations they induce: this phase coherence possibly enables an optimal energy transfer, and hence an exponential growth of regular waves even from very fine seeding wrinkles.

\section{Conclusion}

In this paper a spectral theory is derived to describe the surface deformations of small amplitude under arbitrary normal and tangential stresses applied at the air-liquid interface (wrinkle regime), assuming no feedback of such deformations on the air flow. The key result of the paper is the demonstration of the scaling for the wrinkle amplitude, $\zeta/\delta \simeq (\rho_a / \rho_\ell) u^{* 3/2} / (g \nu_\ell)^{1/2}$, in good agreement with the experimental findings of \cite{Paquier_2015,Paquier_2016}. This theory corresponds to the viscous-limited asymptotic steady state of the inviscid resonant mechanism proposed by \citet{Phillips_1957}, and provides an appropriate description of the surface deformations for wind velocity below the onset of regular waves.

A significant improvement of the present theory is the quantitative description of the fraction of energy supplied by the pressure fluctuations that is located near the resonance. As already pointed out by~\citet{Phillips_1957}, only the pressure fluctuations of space-time correlations matching the dispersion relation  contribute to the surface deformations. Detailed knowledge of the space-time Fourier spectrum of the pressure and shear stress fluctuations in a turbulent boundary layer, which was not available at the time of \citet{Phillips_1957}, was used here to close the problem by determining numerically the dependence of the wrinkle amplitude with the governing parameters. The wrinkle regime therefore provides an interesting configuration where the effect of a turbulent forcing on a dispersive wave system can be exactly computed. \textcolor{black}{A similar approach was recently proposed for waves generated on a viscoelastic compliant coating \citep{Benschop2019}.}

We have shown that the wrinkles below the wave onset correspond to a superposition of uncoherent wakes mostly originating from the pressure fluctuations traveling in the turbulent boundary layer (the shear stress fluctuations are found to provide a negligible contribution to the wrinkles). The thinning of the wrinkles in the spanwise direction as the wind velocity increases is reminiscent of the decrease of the wake angle for a finite-size moving disturbance found in the classical ship wake problem \citep{Rabaud_2013,Darmon_2014,Moisy_2014}.  \textcolor{black}{This mechanism could be related to the surprisingly large cross-wind wave slopes found in ocean observations \citep{Munk2009}.}

A remarkable property of the wrinkle is that their characteristic size are governed by the (outer) boundary layer thickness $\delta$, although they originate from pressure patches of characteristic size governed by the (inner) viscous sublayer thickness, $\delta_\nu \simeq \delta Re_\delta^{-1}$. This is because the liquid surface response integrates the pressure forcing, resulting in a systematic shift towards the upper bound of the energy-containing range $[\delta_\nu, \delta]$ of the forcing. The wrinkles are therefore essentially governed by the largest scales of the turbulent flow. As a consequence, the detailed statistics of the wrinkles is expected to depend on the geometry of the forcing, making a fine comparison between simulations, laboratory and outdoors experiments difficult.

The present theory neglects the influence of surface drift current. Although significant effects of the liquid current are certainly present for the onset and amplification of regular waves, we expect weak influence of drift current on the wrinkle regime. Indeed, the dominant Fourier modes are oriented along the wind direction for regular waves, while they are nearly normal to the wind in the wrinkle regime. As a consequence, only a small Doppler shift will arise in the wrinkle regime, even in the presence of a strong drift current along the wind direction. We therefore expect robust properties of the wrinkles, that may be extended to liquids of small viscosity and large depth, provided that the flow in the liquid remains laminar. As the laminar condition is not satisfied for the air-sea interface, the influence of water current on the wrinkles in oceanographic conditions still deserves further analysis

In spite of this limitation, the implications of the present work for physical oceanography are important. In particular, the transition towards regular (quasi-monochromatic) waves as the wind velocity is increased raises the question of the role of the wrinkles as a base state for wave amplification. The experiments of \cite{Paquier_2016} suggest that the regular waves are triggered when the wrinkle amplitude reaches a fraction of the viscous sublayer thickness. Beyond that amplitude, the feedback of the surface roughness on the turbulent boundary layer can no longer be neglected. This provides a criterion for the wave onset, $u^*_c \propto \nu_\ell^{1/5}$, which is consistent with experiments performed in viscous liquids, and for which no explanation has been proposed so far. Describing the surface deformations for a wind velocity above this threshold is beyond the scope of the present linear theory, in which such coupling between the liquid and the air phases is ignored. This deserves further investigation, as it could renew our understanding on the onset of wave generation.

\section*{Acknowledgements}\label{sec:acknowledgements}

The authors thank L. Deike, C. Garrett, J. Jim\'enez, W. Munk, C. Nov\'e-Josserand, A. Paquier and E. Rapha\"el for fruitful discussions. This work was supported by the  project ``ViscousWindWaves'' (ANR-18-CE30-0003) of the French National Research Agency, and by the LabeX LaSIPS (ANR-10-LABX-0040-LaSIPS) managed by the French National Research Agency under the "Investissements d'avenir" program (ANR-11-IDEX-0003-02). A.L.D. acknowledges the support from the Office of Naval Research under Grant \#N00014-16-S-BA10.

\appendix
\section{Detailed calculation of Eq.~(\ref{eq_perrard})}

We detail here the calculation steps of section 3.3 which establishes the final expression of Eq.~(\ref{eq_perrard}). After introducing the Fourier transforms of $p_\ell, \bf \Omega$ and $\bs v$ and using the boundary conditions in $z=0$ we obtain the system of equation
\bb
\left ( 1 - \frac{2 k^2 \nu_\ell}{i \omega} \right )\frac{k}{\rho_\ell} \hat p_0- \frac{2 \nu_\ell m k  }{i \omega} \nu_\ell \hat B_z -g' k \hat \zeta &=& \frac{k \hat N}{\rho_\ell} \label{eq_nstressF0}\\
\nu_\ell \hat B_z+ 2 \nu_\ell\mathcal{F}\{(\partial_{zz} v_z)_{z=0}\} &=&- \frac{i \bs k \cdot \hat{\bs T}}{\rho_\ell} \ , \label{eq_tstressF0}
\ee
where $g' = g + \gamma k^2/\rho_\ell$ is the modified gravity and $\hat B_z =  (\bs \kappa \times \hat {\bs \Omega}_0(\bs k, \omega)) \cdot \bs e_z$ is the non potential flow part of $\hat v_z$, which satisfies $\Delta v_z = - B_z$. In Fourier space, we can evaluate $\mathcal{F}\{(\partial_{zz} v_{z})_{z=0} \}$ by derivating twice the expression of $\hat v_z$ with respect to $z$,
\bb 
\mathcal{F}\{(\partial_{zz} v_{z})_{z=0} \}= \frac{k^3}{\rho_\ell i \omega} \hat p_0 + \frac{\nu_\ell m^2}{i \omega} \hat B_z\  .
\label{eq_Fzz}
\ee
Using the kinematic condition $\partial_t \zeta = (v_z)_{z=\zeta}$ in the small perturbation limit, we obtain the relation between $\hat \zeta, \hat p_0$ and $\hat B_z$,
\bb
\omega^2 \hat \zeta = \frac{k}{\rho_\ell} \hat p_0 +  \nu_\ell \hat B_z \label{eq_zpBA}\ ,
\ee
Replacing $\mathcal{F}\{(\partial_{zz} v_{z})_{z=0} \}$ by its expression in Eq.~(\ref{eq_tstressF}) yields
\bb
\nu_\ell \hat B_z+ 2 \nu_\ell \left( \frac{k^3}{\rho_\ell i \omega} \hat p_0 + \frac{\nu_\ell m^2}{i \omega} \hat B_z \right) &=&- \frac{i \bs k \cdot \hat{\bs T}}{\rho_\ell}.
\ee
Using the relation $m^2 = k^2 - i \omega/\nu_\ell$, we obtain
\bb
- g' k \hat \zeta +\frac{m^2+k^2}{m^2-k^2}\frac{k}{\rho_\ell} \hat p_0 + \frac{2 m k  }{m^2-k^2} \nu_\ell\hat B_z &=& \frac{k \hat N}{\rho_\ell} \\
\frac{2 k^2  }{m^2-k^2}  \frac{k}{\rho_\ell} \hat p_0 + \frac{m^2+k^2}{m^2-k^2} \nu_\ell\hat B_z &=& \frac{i \bs k \cdot \hat{\bs T}}{\rho_\ell}\ .
\ee
This expression can be simplified by replacing $\hat p_0$ by its expression from Eq.~(\ref{eq_zpBA}) 
\bb
(\omega^2 - g' k) \hat \zeta - \frac{2m}{m+k} \nu_\ell\hat B_z &=& \frac{k \hat N}{\rho_\ell} - \frac{i \bs k \cdot \hat{\bs T}}{\rho_\ell}  \label{eq_nstressF2} \\
\omega^2 \hat \zeta + \frac{m^2-k^2}{2 k^2} \nu_\ell\hat B_z &=& \frac{m^2-k^2}{2k^2}   \frac{i \bs k \cdot \hat{\bs T}}{\rho_\ell} \ . \label{eq_tstressF2}
\ee
Multiplying Eq.~(\ref{eq_tstressF2}) by $(m-k)m/k$ gives
\bb
\frac{4m k^2}{(m+k)(m^2-k^2)}\omega^2 \hat \zeta + \frac{2m}{m+k} \nu_\ell\hat B_z &=& \frac{2m}{m+k}   \frac{i \bs k \cdot \hat{\bs T}}{\rho_\ell} \ .  \label{eq_tstressF3}
\ee
Summing Eqs.~(\ref{eq_nstressF2}) and (\ref{eq_tstressF3}) yields
\bb
\frac{4m k^2}{(m+k)(m^2-k^2)}\omega^2 \hat \zeta+ (\omega^2 - g' k) \hat \zeta = \frac{k \hat N}{\rho_\ell} + \frac{m-k}{m+k}\frac{i \bs k \cdot \hat{\bs T}}{\rho_\ell}.
\ee
Factorizing by $\hat \zeta$, we obtain the final expression 
\bb
\left (\omega^2 - g' k + 4 i \nu_\ell \omega k^2 + 4 \nu_\ell^2 k^3 (m-k) \right) \hat \zeta &=& \frac{k \hat N}{\rho_\ell} + \frac{m-k}{m+k}\frac{i \bs k \cdot \hat{\bs T}}{\rho_\ell},
\ee
which is Eq.~(\ref{eq_perrard}).

\bibliographystyle{jfm}

\bibliography{biblio_WbyW}

\begin{thebibliography}{66}
\expandafter\ifx\csname natexlab\endcsname\relax\def\natexlab#1{#1}\fi

\bibitem[Banner \& Peirson(1998)]{Banner_1998}
{\sc Banner, M.~L. \& Peirson, W.~L.} 1998 Tangential stress beneath
  wind-driven air-water interfaces. {\em J. Fluid Mech.\/} {\bf 364}, 115--145.

\bibitem[Belcher \& Hunt(1998)]{belcher1998turbulent}
{\sc Belcher, SE \& Hunt, JCR} 1998 Turbulent flow over hills and waves. {\em
  Annual Review of Fluid Mechanics\/} {\bf 30}~(1), 507--538.

\bibitem[Benschop {\em et~al.\/}(2019)Benschop, Greidanus, Delfos, Westerweel
  \& Breugem]{Benschop2019}
{\sc Benschop, H.O.G, Greidanus, A.J., Delfos, R., Westerweel, J. \& Breugem,
  W.P.} 2019 Deformation of a linear viscoelastic compliant coating in a
  turbulent flow. {\em J. Fluid Mech.\/} {\bf 859}, 613--658.

\bibitem[Caulliez {\em et~al.\/}(2008)Caulliez, Makin \&
  Kudryavtsev]{Caulliez_2008}
{\sc Caulliez, G., Makin, V. \& Kudryavtsev, V.} 2008 Drag of the water surface
  at very short fetches: Observations and modeling. {\em Journal of Physical
  Oceanography\/} {\bf 38}~(9), 2038--2055.

\bibitem[Choi \& Moin(1990)]{Choi_1990}
{\sc Choi, H. \& Moin, P.} 1990 On the space-time characteristics of
  wall-pressure fluctuations. {\em Physics of Fluids A: Fluid Dynamics
  (1989-1993)\/} {\bf 2}~(8), 1450--1460.

\bibitem[Corcos(1963)]{Corcos_1963}
{\sc Corcos, G.~M.} 1963 The structure of the turbulent pressure field in
  boundary-layer flows. {\em J. Fluid Mech.\/} {\bf 18}.

\bibitem[Darmon {\em et~al.\/}(2014)Darmon, Benzaquen \&
  Rapha\"el]{Darmon_2014}
{\sc Darmon, A., Benzaquen, M. \& Rapha\"el, E.} 2014 Kelvin wake pattern at
  large {F}roude numbers. {\em J. Fluid Mech.\/} {\bf 738}, R3.

\bibitem[Druzhinin {\em et~al.\/}(2012)Druzhinin, Troitskaya \&
  Zilitinkevich]{Druzhinin_2012}
{\sc Druzhinin, O., Troitskaya, A. \& Zilitinkevich, Y.~I.} 2012 Direct
  numerical simulation of a turbulent wind over a wavy water surface. {\em J.
  Geophys. Research\/} {\bf 117}~(C11).

\bibitem[Eckart(1953)]{Eckart_1953}
{\sc Eckart, C.} 1953 The generation of wind waves on a water surface. {\em
  Journal of Applied Physics\/} {\bf 24}~(12), 1485--1494.

\bibitem[Ellingsen \& Li(2017)]{Ellingsen_2017}
{\sc Ellingsen, S.~A. \& Li, Y.} 2017 Approximate dispersion relations for
  waves on arbitrary shear flows. {\em Journal of Geophysical Research:
  Oceans\/} {\bf 122}, 9889--9905.

\bibitem[Francis(1956)]{Francis_1956}
{\sc Francis, J. R.~D.} 1956 {LXIX}. {C}orrespondence. {W}ave motions on a free
  oil surface. {\em Philosophical Magazine\/} {\bf 1}~(7), 685--688.

\bibitem[Funada \& Joseph(2001)]{Funada_2001}
{\sc Funada, T \& Joseph, DD} 2001 Viscous potential flow analysis of
  {K}elvin--{H}elmholtz instability in a channel. {\em J. Fluid Mech.\/} {\bf
  445}, 263--283.

\bibitem[Gottifredi \& Jameson(1970)]{Gottifredi_1970}
{\sc Gottifredi, J. \& Jameson, G.} 1970 The growth of short waves on liquid
  surfaces under the action of a wind. {\em Proceedings of the Royal Society of
  London. A. Mathematical and Physical Sciences\/} {\bf 319}~(1538), 373--397.

\bibitem[Havelock(1919)]{havelock1919}
{\sc Havelock, T.~H.} 1919 Wave resistance: Some cases of three-dimensional
  fluid motion. {\em Proceedings of the Royal Society of London. Series A\/}
  {\bf 95}, 354--365.

\bibitem[Janssen(2004)]{Janssen_2004}
{\sc Janssen, P.} 2004 {\em The interaction of ocean waves and wind\/}.
  Cambridge University Press.

\bibitem[Jimenez(2013)]{Jimenez_2013}
{\sc Jimenez, J.} 2013 Near wall turbulence. {\em Phys. Fluids\/} {\bf 25},
  101302.

\bibitem[Jimenez {\em et~al.\/}(2004)Jimenez, Del~Alamo \&
  Flores]{Jimenez_2004}
{\sc Jimenez, J., Del~Alamo, J.~C. \& Flores, O.} 2004 The large-scale dynamics
  of near-wall turbulence. {\em J. Fluid Mech.\/} {\bf 505}, 179--199.

\bibitem[Jimenez \& Hoyas(2008)]{Jimenez_2008}
{\sc Jimenez, J. \& Hoyas, S.} 2008 Turbulent fluctuations above the buffer
  layer of wall-bounded flows. {\em J. Fluid Mech.\/} {\bf 611}, 215--236.

\bibitem[Jimenez {\em et~al.\/}(2010)Jimenez, Hoyas, Simens \&
  Mizuno]{Jimenez_2010}
{\sc Jimenez, J., Hoyas, S., Simens, M.~P. \& Mizuno, Y.} 2010 Turbulent
  boundary layers and channels at moderate {R}eynolds numbers. {\em J. Fluid
  Mech.\/} {\bf 657}, 335--360.

\bibitem[Kahma \& Donelan(1988)]{Kahma_1988}
{\sc Kahma, K. \& Donelan, M.~A.} 1988 A laboratory study of the minimum wind
  speed for wind wave generation. {\em J. Fluid Mech.\/} {\bf 192}, 339--364.

\bibitem[Kawai(1979)]{Kawai_1979}
{\sc Kawai, S.} 1979 Generation of initial wavelets by instability of a coupled
  shear flow and their evolution to wind waves. {\em J. Fluid Mech.\/} {\bf
  93}~(4), 661--703.

\bibitem[Keulegan(1951)]{Keulegan_1951}
{\sc Keulegan, G.~H.} 1951 Wind tides in small closed channels. {\em Journal of
  Research of the National Bureau of Standards\/} {\bf 46}, 358--381.

\bibitem[Kim {\em et~al.\/}(2011)Kim, Padrino \& Joseph]{Kim_2011}
{\sc Kim, H, Padrino, J.~C. \& Joseph, D.~D.} 2011 Viscous effects on
  {K}elvin--{H}elmholtz instability in a channel. {\em J. Fluid Mech.\/} {\bf
  680}, 398--416.

\bibitem[Kim(1989)]{Kim1989}
{\sc Kim, J.} 1989 On the structure of pressure fluctuations in simulated
  turbulent channel flow. {\em J. Fluid Mech.\/} {\bf 205}, 421--451.

\bibitem[Kim {\em et~al.\/}(1987)Kim, Moin \& Moser]{kim:moi:mos:87}
{\sc Kim, J., Moin, P. \& Moser, R.~D.} 1987 Turbulence statistics in fully
  developed channel flow at low {R}eynolds number. {\em J. Fluid Mech.\/} {\bf
  177}, 133--166.

\bibitem[Kirby \& Chen(1989)]{Kirby_1990}
{\sc Kirby, J.~T. \& Chen, T.~M.} 1989 Surface waves on vertically sheared
  flows: Approximate dispersion relations. {\em J. Geophys. Research Oceans\/}
  {\bf 94}.

\bibitem[Kudryavtsev {\em et~al.\/}(2014)Kudryavtsev, Chapron \&
  Makin]{Kudryavtsev_2014}
{\sc Kudryavtsev, V, Chapron, B. \& Makin, V} 2014 Impact of wind waves on the
  air-sea fluxes: A coupled model. {\em Journal of Geophysical Research:
  Oceans\/} {\bf 119}~(2), 1217--1236.

\bibitem[Kudryavtsev \& Makin(2002)]{Kudryavtsev_2002}
{\sc Kudryavtsev, V.~N. \& Makin, V.~K.} 2002 Coupled dynamics of short waves
  and the airflow over long surface waves. {\em J. Geophys. Research\/} {\bf
  107}~(C12), 3209.

\bibitem[Lamb(1995)]{Lamb_1995}
{\sc Lamb, H.} 1995 {\em Hydrodynamics\/}. Sixth edition, Cambridge University
  Press.

\bibitem[Langevin(1908)]{Langevin_1908}
{\sc Langevin, P.} 1908 Sur la th\'eorie du mouvement brownien. {\em C. R.
  Acad. Sci.\/} {\bf 146}, 530--533.

\bibitem[LeBlond \& Mainardi(1987)]{Leblond_1987}
{\sc LeBlond, P.H. \& Mainardi, F.} 1987 The viscous damping of
  capillary-gravity waves. {\em Acta Mechanica\/} {\bf 68}, 203--222.

\bibitem[Lee \& Moser(2015)]{Lee_JFM_2015}
{\sc Lee, M. \& Moser, R.~D.} 2015 Direct numerical simulation of a turbulent
  channel flow up to ${R}e_\tau \approx 5200$. {\em J. Fluid Mech.\/} {\bf
  774}, 395--415.

\bibitem[Liberzon \& Shemer(2011)]{Liberzon_2011}
{\sc Liberzon, D. \& Shemer, L.} 2011 Experimental study of the initial stages
  of wind waves' spatial evolution. {\em J. Fluid Mech.\/} {\bf 681}, 462--498.

\bibitem[Lin {\em et~al.\/}(2008)Lin, Moeng, Tsai, Sullivan \&
  Belcher]{Lin_2008}
{\sc Lin, M.-Y., Moeng, C.-H., Tsai, W.-T., Sullivan, P.~P. \& Belcher, S.~E.}
  2008 Direct numerical simulation of wind-wave generation processes. {\em J.
  Fluid Mech.\/} {\bf 616}, 1--30.

\bibitem[Lindsay(1984)]{Lindsay_1984}
{\sc Lindsay, K.~A.} 1984 The {K}elvin-{H}elmholtz instability for a viscous
  interface. {\em Acta mechanica\/} {\bf 52}~(1), 51--61.

\bibitem[Lozano-Dur{\'a}n \& Jim{\'e}nez(2014)]{lozano2014time}
{\sc Lozano-Dur{\'a}n, A. \& Jim{\'e}nez, J.} 2014 Time-resolved evolution of
  coherent structures in turbulent channels: characterization of eddies and
  cascades. {\em J. Fluid Mech.\/} {\bf 759}, 432--471.

\bibitem[Manneville(2010)]{manneville2010instabilities}
{\sc Manneville, P.} 2010 {\em Instabilities, {C}haos and {T}urbulence\/}.
  World Scientific.

\bibitem[Melville {\em et~al.\/}(1998)Melville, Shear \& Veron]{Melville_1998}
{\sc Melville, W.~K., Shear, R. \& Veron, F.} 1998 Laboratory measurements of
  the generation and evolution of langmuir circulations. {\em J. Fluid Mech.\/}
  {\bf 364}, 31--58.

\bibitem[Miles(1957)]{Miles_1957}
{\sc Miles, J.~W.} 1957 On the generation of surface waves by shear flows. {\em
  J. Fluid Mech.\/} {\bf 3}, 185--204.

\bibitem[Miles(1968)]{Miles_1968}
{\sc Miles, J.~W.} 1968 The {C}auchy--{P}oisson problem for a viscous liquid.
  {\em J. Fluid Mech.\/} {\bf 34}~(02), 359--370.

\bibitem[Miles(1993)]{Miles_1993}
{\sc Miles, J.~W.} 1993 Surface-wave generation revisited. {\em J. Fluid
  Mech.\/} {\bf 256}, 427--441.

\bibitem[Moisy \& Rabaud(2014{\natexlab{{\em a\/}}})]{Moisy_PRE_2014}
{\sc Moisy, F. \& Rabaud, M.} 2014{\natexlab{{\em a\/}}} Mach-like
  capillary-gravity wakes. {\em Phys. Rev. E\/} {\bf 90}, 023009.

\bibitem[Moisy \& Rabaud(2014{\natexlab{{\em b\/}}})]{Moisy_2014}
{\sc Moisy, F. \& Rabaud, M.} 2014{\natexlab{{\em b\/}}} Scaling of far-field
  wake angle of non-axisymmetric pressure disturbance. {\em Phys. Rev. E\/}
  {\bf 89}, 063004.

\bibitem[Moisy {\em et~al.\/}(2009)Moisy, Rabaud \& Salsac]{Moisy_2009}
{\sc Moisy, F., Rabaud, M. \& Salsac, K.} 2009 A synthetic schlieren method for
  the measurement of the topography of a liquid interface. {\em Exp. Fluids\/}
  {\bf 46}, 1021--1036.

\bibitem[Moser {\em et~al.\/}(1999)Moser, Kim \& Mansour]{Moser_1999}
{\sc Moser, R.~D., Kim, J. \& Mansour, N.~N.} 1999 Direct numerical simulation
  of turbulent channel flow up to {R}e = 590. {\em Phys. Fluids\/} {\bf
  11}~(4), 943--945.

\bibitem[Munk(2009)]{Munk2009}
{\sc Munk, W.} 2009 An inconvenient sea truth: Spread, steepness, and skewness
  of surface slopes. {\em Annu. Rev. Mar. Sci\/} {\bf 1}, 377--415.

\bibitem[Paquier {\em et~al.\/}(2015)Paquier, Moisy \& Rabaud]{Paquier_2015}
{\sc Paquier, A., Moisy, F. \& Rabaud, M.} 2015 Surface deformations and wave
  generation by wind blowing over a viscous liquid. {\em Phys. Fluids\/} {\bf
  27}, 122103.

\bibitem[Paquier {\em et~al.\/}(2016)Paquier, Moisy \& Rabaud]{Paquier_2016}
{\sc Paquier, A., Moisy, F. \& Rabaud, M.} 2016 Viscosity effects in wind wave
  generation. {\em Phys. Rev. Fluids\/} {\bf 1}, 083901.

\bibitem[Peregrine(1976)]{Peregrine_1976}
{\sc Peregrine, D.~H.} 1976 Interaction of water waves and currents. {\em
  Advances in Applied Mechanics.\/} {\bf 16}, 9--117.

\bibitem[Phillips(1957)]{Phillips_1957}
{\sc Phillips, O.~M.} 1957 On the generation of waves by turbulent wind. {\em
  J. Fluid Mech.\/} {\bf 2}~(05), 417--445.

\bibitem[Plant(1982)]{Plant_1982}
{\sc Plant, W.~J.} 1982 A relationship between wind stress and wave slope. {\em
  Journal of Geophysical Research: Oceans (1978--2012)\/} {\bf 87}~(C3),
  1961--1967.

\bibitem[Pottier(2014)]{Pottier_2014}
{\sc Pottier, N.} 2014 {\em Non Equilibrium Statistical Physics\/}. Oxford
  Graduate texts.

\bibitem[Rabaud \& Moisy(2013)]{Rabaud_2013}
{\sc Rabaud, M. \& Moisy, F.} 2013 Ship wakes: {K}elvin or {M}ach angle? {\em
  Phys. Rev. Lett.\/} {\bf 110}, 214503.

\bibitem[Rapha{\"e}l \& de~Gennes(1996)]{Raphael_1996}
{\sc Rapha{\"e}l, E. \& de~Gennes, P-G.} 1996 Capillary gravity waves caused by
  a moving disturbance: wave resistance. {\em Physical Review E\/} {\bf
  53}~(4), 3448.

\bibitem[Richard \& Rapha\"el(1999)]{Richard_1999}
{\sc Richard, D. \& Rapha\"el, E.} 1999 Capillary-gravity waves: The effect of
  viscosity on the wave resistance. {\em EPL (Europhysics Letters)\/} {\bf
  48}~(1), 49.

\bibitem[Robinson(1991)]{Robinson_1991}
{\sc Robinson, S.~K.} 1991 Coherent motions in the turbulent boundary layer.
  {\em Annu. Rev. Fluid Mech.\/} {\bf 23}, 601--639.

\bibitem[Russell(1844)]{Russell_1844}
{\sc Russell, J.~S.} 1844 On waves. In {\em Report of fourteenth meeting of the
  British Association for the Advancement of Science, York\/}, pp. 311--390.

\bibitem[Sajjadi {\em et~al.\/}(2017)Sajjadi, Robertson, Harvey \&
  Brown]{sajjadi2017wave}
{\sc Sajjadi, S.~G., Robertson, S., Harvey, R. \& Brown, M.} 2017 Wave motion
  induced by turbulent shear flows over growing {S}tokes waves. {\em Journal of
  Ocean Engineering and Marine Energy\/} {\bf 3}~(2), 97--112.

\bibitem[Schlichting(2000)]{Schlichting_2000}
{\sc Schlichting, H.} 2000 {\em Boundary Layer Theory\/}, 8th edn. Springer.

\bibitem[Sullivan \& McWilliams(2010)]{Sullivan_2010}
{\sc Sullivan, P.~P. \& McWilliams, J.~C.} 2010 Dynamics of winds and currents
  coupled to surface waves. {\em Annu. Rev. Fluid Mech.\/} {\bf 42}, 19--42.

\bibitem[Veron \& Melville(2001)]{Veron_2001}
{\sc Veron, F. \& Melville, W.~K.} 2001 Experiments on the stability and
  transition of wind-driven water surfaces. {\em J. Fluid Mech.\/} {\bf
  446}~(10), 25--65.

\bibitem[Willmarth \& Wooldridge(1962)]{Willmarth_1962}
{\sc Willmarth, W.~W. \& Wooldridge, C.~E.} 1962 Measurements of the
  fluctuating pressure at the wall beneath a thick turbulent boundary layer.
  {\em J. Fluid Mech.\/} {\bf 14}, 187--210.

\bibitem[Yamamoto \& Tsuji(2018)]{Yamamoto_PRF_2018}
{\sc Yamamoto, Y. \& Tsuji, Y.} 2018 Numerical evidence of logarithmic regions
  in channel flow at ${R}e_\tau = 8000$. {\em Phys. Rev. Fluids\/} {\bf
  3}~(012602(R)).

\bibitem[Zavadasky \& Shemer(2017)]{Zavadasky_2017}
{\sc Zavadasky, A. \& Shemer, L.} 2017 Water waves excited by near-impulsive
  wind forcing. {\em J. Fluid Mech.\/} {\bf 828}, 459--495.

\bibitem[Zhang(1995)]{Zhang_1995}
{\sc Zhang, X.} 1995 Capillary--gravity and capillary waves generated in a wind
  wave tank: Observations and theories. {\em J. Fluid Mech.\/} {\bf 289},
  51--82.

\bibitem[Zonta {\em et~al.\/}(2015)Zonta, Soldati \& Onorato]{Zonta_2015}
{\sc Zonta, F., Soldati, A. \& Onorato, M.} 2015 Growth and spectra of
  gravity--capillary waves in countercurrent air/water turbulent flow. {\em J.
  Fluid Mech.\/} {\bf 777}, 245--259.

\end{thebibliography}

\end{document}